\documentclass[
nofootinbib,
 amsmath,amssymb,
 aps,
pra,
onecolumn
]{revtex4-2}

\usepackage{float}
\usepackage{algorithm}
\usepackage{algpseudocode}
\usepackage{graphicx}
\usepackage{subfigure}
\usepackage{amssymb,amsfonts,amsmath,amsthm}
\usepackage{braket}
\usepackage{hyperref}
\usepackage{color}
\usepackage{soul}
\usepackage{changes}

\usepackage{enumerate,enumitem}
\usepackage{comment}
\usepackage{optidef}
\usepackage{booktabs}

\usepackage{quantikz}
\usepackage{tikz}
\usetikzlibrary{tikzmark,calc}

\newtheorem{proposition}{Proposition}

\newtheorem{theorem}{Theorem}
\newtheorem{remark}{Remark}

\newtheorem{lemma}{Lemma}

\def\R{\mathbb{R}}

\newcommand{\IFLP}{ Instituto de F\'isica La Plata (CONICET-UNLP), La Plata,
Buenos Aires, Argentina}

\newcommand{\ICC}{
Instituto de Ciencias de la Computaci\'on, CONICET - Universidad de Buenos Aires, Buenos Aires, Argentina}

\newcommand{\CITECCA}{Centro Interdisciplinario de Telecomunicaciones Electr\'onica, Computaci\'on y Ciencia Aplicada, Universidad Nacional de R\'io Negro, CONICET, Bariloche, R\'io Negro, Argentina}

\usepackage{xcolor}
\begin{document}

\title{Quantum sine imaginary-time evolution: sequential and QSVT implementations of a probabilistic ground-state filter}%

\author{Victor A. Penas}
\email{vapenas@unrn.edu.ar}
\affiliation{\CITECCA}

\author{Tamara Bottazzi}
\affiliation{\CITECCA}

\author{Marcelo Losada}
\affiliation{\ICC}

\author{Diego Tielas}
\affiliation{\IFLP}

\author{Federico Holik}
\affiliation{Universidad Nacional de Hurlingham (UNAHUR), Laboratorio de Investigaci\'on y Desarrollo Experimental en Computaci\'on (LIDEC), Hurlingham, Argentina}

\date{\today}

\begin{abstract}
	We propose quantum sine imaginary-time evolution (QSITE), a quantum algorithm for preparing ground states of a Hamiltonian. QSITE approximates imaginary-time evolution through an exact sine filter. We give two circuit implementations of this operator: a sequential one that applies it as a concatenation of linear combinations of unitaries (LCU), reusing two ancilla qubits across steps via mid-circuit measurement, and a coherent one based on the quantum singular value transformation (QSVT). A telescoping identity shows the two prepare the same state and succeed with identical probability, differing only in how that probability can be boosted.
QSITE belongs to the family of ancilla-based probabilistic imaginary-time filters and we state rigorous convergence guarantees: we prove that the QSITE operator converges uniformly to the exact ITE operator, that the prepared state converges to the ideal ITE state, and that the filter probability increases monotonically with each concatenation step.
Although the construction applies to an arbitrary Hamiltonian, we specialize it to optimization Hamiltonians, and this specialization is what exposes its structure more easily.
Since MaxCut eigenvalues are integers, the probability of failure is periodic in the evolution time, yielding safe zones, the first located analytically and the second numerically. For MaxCut we further show that, when the tolerated energy error is a fixed fraction of the ground-state energy, the probability of failure can be bounded below $1/2$ with concatenation steps growing linearly with the number of qubits in the first safe zone. We present numerical results for random Erd\"os--R\'enyi instances with up to $24$ qubits and benchmark QSITE against QAOA. 

\end{abstract}

\maketitle

\section{Introduction}
Determining the ground state of a Hamiltonian is a central task in many areas of physics and is crucial for solving a wide range of optimization problems. Quantum computing offers promising pathways to address this challenge. 
A variety of quantum algorithms have been developed for this purpose. Some are general-purpose schemes that apply to an arbitrary Hamiltonian, such as quantum phase estimation \cite{Kitaev_1995, Abrams_1999}, quantum Krylov subspace methods \cite{Klymko_2022}, the rodeo algorithm \cite{Choi_2021_Rodeo}, ground-state filtering methods built from the real-time evolution operator, such as that of Ge, Tura and Cirac \cite{GeTuraCirac_2019}, the quantum eigenvalue transformation of unitary matrices (QETU) \cite{Dong_2022_QETU}, which has been demonstrated on trapped-ion hardware with an adiabatically prepared input state \cite{Karacan_2025}, and power-cosine filtering \cite{Jo_2026}, and the Variational Quantum Eigensolver (VQE) \cite{Peruzzo_2014, Cerezo_2021, Tilly_2022}. Others have been developed specifically for optimization problems, including adiabatic quantum optimization \cite{Farhi_2001}, quantum annealing \cite{Das_2008}, and the Quantum Approximate Optimization Algorithm (QAOA) \cite{farhi2014quantumapproximateoptimizationalgorithm,Zhou_2020,ostrowski2020lowerboundscircuitdepth,Shaydulin_2021,Medvidovi__2021,PhysRevResearch.4.033029,herrman2021multianglequantumapproximateoptimization}. 

Despite significant progress, these methods face potential limitations, particularly on near-term quantum devices with constrained resources. For instance, adiabatic quantum optimization may offer only polynomial speedups for certain problems \cite{Lucas:2013ahy, Abbas_2024}. Variational quantum algorithms, on the other hand, are often constrained by the expressiveness of a fixed ansatz and by the difficulty of the associated high-dimensional classical optimization, whose cost-function landscapes can exhibit exponentially vanishing gradients, the so-called barren plateaus \cite{McClean_2018}. These obstacles motivate the search for ground-state preparation schemes that reduce or eliminate the reliance on classical parameter optimization.

An alternative family of quantum algorithms for finding quantum ground states for general Hamiltonians, grouped under the umbrella term quantum imaginary-time evolution (QITE), tries to implement the non-unitary imaginary-time evolution operator (ITE), $e^{-tH}$, by using unitary operators and different techniques. The ITE operator drives the quantum system towards its ground state as time proceeds by effectively cooling it to zero temperature \cite{Love_2020}. Motta et al \cite{Motta_2019} implements QITE by using measurement-assisted unitary operators in each step of a Trotter decomposition and choosing a domain size for the expansion of these operators in terms of Pauli strings. Several enhancements of this protocol target NISQ devices \cite{Nishi_2021, Ville_2022, Cao_2022, Benedetti_2021, Gacon_2023, Gacon_2024}, and further variants rely on Monte Carlo sampling \cite{Huo_2023}, reinforcement learning \cite{Cao_2022}, or an orthogonal basis for each propagated state \cite{Jouzdani_2022}. A variational implementation, known as variational imaginary-time evolution (VITE) \cite{Jones_2019, McArdle_2019, Yuan_2019, Beach_2019, Gacon_2023, Gacon_2024, Kolotouros_2025}, instead solves the imaginary-time Schr\"odinger equation on a classical computer for the parameters of unitary ansatz gates, the governing equation being derived from McLachlan's variational principle \cite{McLachlan_1964}; being a variant of the variational quantum eigensolver (VQE), the ansatz must be selected carefully to reach a state close to the true ground state. A third route, probabilistic imaginary-time evolution (PITE), realizes the non-unitary operator directly through measurement: the system register is evolved unitarily on an enlarged space together with an ancilla, and a subsequent measurement of the ancilla collapses the system onto the (normalized) ITE-evolved state whenever the successful outcome is obtained \cite{Lin_2021, Liu_2021}. Unlike VITE and QITE, PITE introduces neither an ansatz nor a locality-based truncation, so the energy is minimized over the full Hilbert space of the register; its price is a success probability that decays as the imaginary time grows. Kosugi et al.\ implement PITE with a single ancilla, approximating the non-unitary step by a linear combination of the forward and backward real-time evolutions $e^{\pm iH\Delta\tau}$ and thereby avoiding any eigendecomposition of $H$ \cite{PhysRevResearch.4.033121}; subsequent work restores an efficient success probability through quantum amplitude amplification---rendering the evolution deterministic in the ideal case---\cite{Nishi_2022_QAA} and analyzes the optimal scheduling of the imaginary-time step \cite{Nishi_2023_scheduling}. 

QITE has been successfully implemented on noisy intermediate-scale quantum (NISQ) hardware for applications in quantum chemistry \cite{Barison_2022, Yeter-Aydeniz_2020, Tsuchimochi_2023}, open quantum system simulations \cite{Kamakari_2022}, and the calculation of thermodynamic properties \cite{Getelina_2023, Gacon_2024}. More recently, its application has expanded to optimization problems such as Max-Cut, polynomial unconstrained Boolean optimization (PUBO), and the unit-disk maximum independent set (UD-MIS) problem \cite{alam2023solvingmaxcutquantumimaginary, PhysRevA.109.052430, PhysRevA.111.042403}. In particular, PITE has found applications in quantum chemistry, where it was introduced as a first-quantized eigensolver that also prepares finite-temperature Gibbs states and partition functions \cite{PhysRevResearch.4.033121}, and was subsequently applied to molecular ground states with explicit gate-level circuits validated on superconducting and trapped-ion hardware \cite{Xie_2024_PITE}, as well as to condensed-matter spin systems such as the Heisenberg and transverse-field Ising chains, with a trapped-ion demonstration and an accompanying software package \cite{2s2p-kvcx, svPITE_2026}. Recently, PITE has also been directed at combinatorial optimization \cite{Nishi_2022_QAA}, while closely related imaginary-time block encodings  have been benchmarked on Max-Cut in \cite{Zhong_2025_ITEBE} and extended to PUBO and higher-order unconstrained binary optimization \cite{FinITE_2026}, and a probabilistic double-bracket variant of QITE has been applied to weighted Max-Cut starting from QAOA states~\cite{Kolotouros_2026_PDBQITE}.

In this paper we propose the quantum sine imaginary-time evolution (QSITE), a new implementation of imaginary-time evolution. Our starting point is the  observation that the ITE operator $e^{-tH}$ is well approximated by $\left(I-\sin(tH/m)\right)^m$ for large $m$, a quantity that admits an exact decomposition as a concatenation of linear combinations of unitaries (LCU) built from evolutions $e^{\pm i tH/m}$. In the body of the paper we decouple the step size from the number of steps and work with the more general operator $\left(I-\sin(tH/k)\right)^m$; the convergence theorems of Sec.~\ref{sec:qsite} correspond to $k=m$. The resulting circuit has a fixed structure, with no variational parameters. We present two implementations: a sequential one, which uses only two ancilla qubits and reuses them across steps through mid-circuit measurement, and a coherent one based on QSVT \cite{Low_2019_Qubitization, Gilyen_2019_QSVT, Martyn_2021_GrandUnification, Lloyd_2021_HamiltonianQSVT}, which implements the QSITE operator without mid-circuit measurements (exactly, after real-part extraction with one additional ancilla), as we discuss in Sec.~\ref{sec:qsvt-impl}. 

Our main contributions are as follows. (i)~We prove two convergence theorems showing that the QSITE operator converges uniformly to the exact ITE operator and that the prepared state converges to the ideal ITE state as the number of concatenation steps $m$ grows. We also prove, for an arbitrary Hamiltonian, that the probability of passing the ancilla filter increases monotonically from step to step, even though the probability of passing all $m$ filters, being a product of these per-step probabilities, decreases with $m$. (ii)~Specializing to MaxCut, we prove that $m=O(N)$ concatenation steps suffice to bound the probability of failure below $1/2$ when the tolerated energy error is a fixed fraction of the ground-state energy. (iii)~We carry out extensive numerical simulations on random Erd\"os--R\'enyi MaxCut instances with up to $24$ qubits, identify two ``safe zones'' of evolution time in which the probability of failure is controlled, and characterize how the required number of shots grows with system size over the studied range $N=6$--$24$. (iv)~We give two implementations of the QSITE operator, a sequential one and a coherent one based on QSVT, and show through a telescoping identity that they prepare the same state with identical success probability, so that they differ in their resources and, above all, in their compatibility with amplitude amplification. (v)~We compare QSITE with QAOA on identical instances at budgets of circuit executions no larger than those of QAOA, counting the circuit runs QAOA spends on classical optimization against QSITE's measurement shots. 

The remainder of this paper is organized as follows. Section~\ref{sec:qsite} fixes notation, introduces the sine expansion together with the two main convergence theorems, and describes the QSITE quantum circuit. Section~\ref{sec:implementations} presents the sequential and coherent (QSVT) implementations of the QSITE operator, which prepare the same state with the same success probability but differ in their compatibility with amplitude amplification. Section~\ref{sec:optimization-based-on-qite} specializes the method to optimization (MaxCut) Hamiltonians, defines the probability of failure and the relevant filter probabilities, and analyzes the computational complexity. Section~\ref{sec:numerical-results} presents the numerical results, including the safe-zone analysis, amplitude amplification with the QSVT implementation, the shot-scaling study, and the comparison with QAOA. Section~\ref{sec:conclusions} summarizes our findings and outlines open directions. Technical lemmas, proofs, and additional figures are collected in the appendices.

\section{QSITE}\label{sec:qsite}
\subsection{Notation and main Theorems}
The imaginary time evolution method allows one to obtain the ground state $\ket{\psi}$ of the Hamiltonian as the long-time limit of the imaginary time Schr\"odinger equation, that is, 
 
 \begin{equation}\label{eq:ite-state}
 |\psi\rangle = \lim_{t \rightarrow \infty} |\psi^{ite}_t\rangle, ~~~ |\psi^{ite}_t\rangle = \frac{e^{-tH}|\psi_0\rangle} {||e^{-tH}\ket{\psi_0}|| },
 \end{equation}
 for some initial state $|\psi_0\rangle$ ($\langle \psi_0|\psi\rangle \neq 0$). State $|\psi^{ite}_t\rangle$ will be referred to as ITE state. 
The parameter $t$ can also be thought of as the inverse temperature, i.e. $t = \frac{1}{k_B T}$, where $k_B$ is the Boltzmann constant and $T$ is the temperature. Eq. \eqref{eq:ite-state} then states that the system tends to the ground state when temperature goes to zero.

Here, we propose a  new implementation of QITE motivated by the standard approximation formula  of the ITE operator:
\begin{equation}\label{eq:StandExpExpansion}
    e^{-Ht}=\left( I-\frac{Ht}{k}\right)^k + O \Big( \frac{H^2}{k}\Big).
\end{equation}

The idea is that under sufficient conditions we can express the Euler operator $(1-Ht/k)^k$ as a sine expansion:

\begin{align}
    \left( I-\frac{Ht}{k}\right)^k  &\approx \left( I-\sin\Big(\frac{Ht}{k}\Big)\right)^k  \\ \label{eq:sine-concat}
     &\approx  \Big(I -\frac{i}{2}e^{-i\frac{Ht}{k}} + \frac{i}{2}e^{i\frac{Ht}{k}}  \Big)^k ,
\end{align}
where in the last term the usual formula for writing sine in terms of exponentials has been used. This last term suggests that we can implement expansion (\ref{eq:StandExpExpansion}) as a concatenation of an LCU (linear combination of unitaries) as we will see in the next section. We present here our two main theorems that guarantee convergence of our proposed method. Let $d=2^N$ be the dimension of the Hilbert space and $N$ the number of qubits. Given some initial state $|\psi_0\rangle$ we define the QSITE state, $|\psi^{qsite}_{t,k}\rangle$, as: 

\begin{equation} \label{eq:site-state}
    \ket{\psi^{qsite}_{t, k}}= \frac{\left(I+ \dfrac{e^{i\frac{-tH}{k}}-e^{i\frac{tH}{k}}}{2i}\right)^k|\psi^0\rangle}{\left\|\left(I+ \dfrac{e^{i\frac{-tH}{k}}-e^{i\frac{tH}{k}}}{2i}\right)^k|\psi^0\rangle\right\|}. 
\end{equation}
Note that $\left(I+ \dfrac{e^{i\frac{-tH}{k}}-e^{i\frac{tH}{k}}}{2i}\right)^k|\psi^0\rangle\neq 0$ for $k$ sufficiently large (see Proposition \ref{coro2}).

\begin{theorem}\label{teo:site-operator}
    Let $H$ be a Hermitian matrix of $d\times d$, $t\in \R$ and $k\in\mathbb{N}$. Then,
    \begin{equation}\label{conv I-sen(H/k)}
    \lim_{k\to \infty}\left(I+ \dfrac{e^{i\frac{-tH}{k}}-e^{i\frac{tH}{k}}}{2i}\right)^k=e^{-tH}
    \end{equation}   
    Moreover, the convergence is uniform in any interval $[0,c]$, with $c>0$.
\end{theorem}

\begin{theorem} \label{teo:site-state}
    Let $\ket{\psi^{ite}_{t}}$ and $\ket{\psi^{qsite}_{t,k}}$ be the states defined in Eq. \eqref{eq:ite-state} and Eq. \eqref{eq:site-state}, respectively. Then, 
\begin{equation}  \label{lim ite -site}
\lim_{k\to \infty}  \langle\psi^{ite}_{t}|\psi^{qsite}_{t,k}\rangle=1 
\end{equation}
and
\begin{equation}
\lim_{k\to \infty}\left\|\ket{\psi^{ite}_{t}}-\ket{\psi^{qsite}_{t,k}}\right\| =0 
\end{equation}
\end{theorem}

In all previous equations the subdivision parameter $k$ appears also as an exponent. For convenience, we will treat them separately since the exponent will be thought of as a concatenation step.
We thus define the QSITE operator with subdivision $k$ and $m$ concatenations as 

    \begin{equation}\label{eq:site-op-k}
	    S^m_{k}(t)=\left(I-\frac{ie^{-it\frac{H}{k}}}{2}+\frac{ie^{it\frac{H}{k}}}{2}\right)^m = \left(I - \sin\!\left(\frac{tH}{k}\right) \right)^m.
    \end{equation}

Note that the above operator is a well-defined operator for every $t\in\mathbb{R}$, $k\in\mathbb{N}$ and $m\in\mathbb{N}$ regardless of its convergence to the ITE operator. It can be constructed by multiplying $m$ times the single step operator:
\begin{equation}\label{eq:site-lcu}
	S_{k}(t) = I - \sin(tH/k) = I + \tfrac{1}{2}\bigl(-i\,e^{-itH/k}\bigr) + \tfrac{1}{2}\bigl(i\,e^{itH/k}\bigr).
\end{equation}
Eq. (\ref{eq:site-op-k}) can then be thought of as coming from a process of concatenating $m$ steps of Eq. (\ref{eq:site-lcu}) or from a coherent implementation as we will see in Section \ref{sec:implementations}.

\subsection{QSITE circuit}

\begin{figure}[H]
    \centering
    \includegraphics[width=0.62\linewidth]{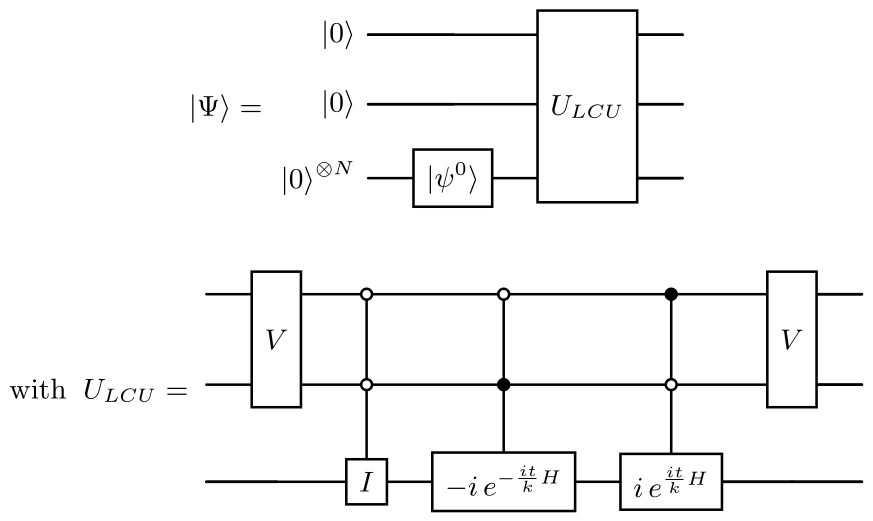}
	  \caption{Quantum circuit to implement the QSITE state (one concatenation) as an LCU.} 
    \label{fig:qsite-circuit-1}
\end{figure}

In order to realize state \eqref{eq:site-state} we propose the  quantum circuit of Fig. (\ref{fig:qsite-circuit-1}).

Where $V$ is a householder transformation given by:
\begin{equation}
   V = I-2|v\rangle\langle v|,
\end{equation}
\begin{equation}
  |v\rangle = \frac{|00\rangle-|\beta\rangle}{{\sqrt{2-2\langle00|\beta\rangle}}}, ~~~~ |\beta\rangle = \frac{|00\rangle+\frac{1}{\sqrt{2}}|01\rangle + \frac{1}{\sqrt{2}}|10\rangle}{\sqrt{2}}. 
\end{equation}
Then, $|\Psi\rangle$ turns out to be:

\begin{align}\nonumber
    |\Psi\rangle &= \frac{|00\rangle}{2}\left[I -\frac{ie^{-\frac{it}{k}H}}{2}+\frac{ie^{\frac{it}{k}H}}{2} \right]|\psi^0\rangle ~~+\\\nonumber
    & \frac{|01\rangle}{2}\left[\frac{I}{\sqrt{2}} -(2-a)\frac{ie^{-\frac{it}{k}H}}{2}-a\frac{ie^{\frac{it}{k}H}}{2} \right]|\psi^0\rangle ~~+\\
    & \frac{|10\rangle}{2}\left[\frac{I}{\sqrt{2}} +a\frac{ie^{-\frac{it}{k}H}}{2}+(2-a)\frac{ie^{\frac{it}{k}H}}{2} \right]|\psi^0\rangle
\end{align}

Here $|\psi_0\rangle$ some initial state and $a=\frac{\sqrt{2}}{2(\sqrt{2}-1)}$.  We can write as $|\Psi\rangle=|00\rangle\otimes|\phi_1\rangle + |\Phi^{\perp}\rangle$ with $|\phi_1\rangle$ and $|\Phi^{\perp}\rangle$ not necessarily normalized. Then:

\begin{align}\nonumber
    |\Psi\rangle &=  |00\rangle\otimes|\phi_1\rangle + |\Phi^\perp\rangle\\
     &=  \sin(\theta_0)|00\rangle\otimes|\tilde{\phi_1}\rangle + \cos(\theta_0)|\tilde{\Phi}^\perp\rangle
\end{align}

with $||\tilde{\phi_1}||=1$, $||\tilde{\Phi}^\perp||=1$ $(\theta\in[0,\pi/2])$. It is easy to see that:
\begin{equation}\label{eq:sin-theta0}
	\sin(\theta_0) = \frac{\Bigg|\Bigg|\left[I -\frac{ie^{-\frac{it}{k}H}}{2}+\frac{ie^{\frac{it}{k}H}}{2} \right]|\psi^0\rangle \Bigg|\Bigg|}{2}=\frac{\Big|\Big|S_k(t)|\psi^0\rangle \Big|\Big|}{2}.
\end{equation}

Measuring the component $|00\rangle$ at the end of circuit Fig. (\ref{fig:qsite-circuit-1}) will return the QSITE state with exponent $m=1$. The probability of measuring the $|00\rangle$ component is equal to $\sin^2(\theta_0)$. Concatenating $m$ blocks $U_{LCU}$, with a measurement after each one, succeeds only if every measurement returns $|00\rangle$. The probability of this event is the product of the $m$ per-step probabilities, and it is therefore non-increasing in $m$, since each factor is at most $1$. The following proposition shows that the per-step probabilities never decrease along the concatenation, so the product is bounded below by $\sin^{2m}(\theta_0)$ and decays at most geometrically.

\begin{proposition}\label{eq:monotoniafiltro}
	Let $|\psi^0\rangle\in \mathcal{H} $ be an initial state $\|\psi^0\|=1$, $H$ a Hermitian operator acting on $\mathcal{H}$ and $S_k$ defined in Eq. \eqref{eq:site-lcu} with $k\in\mathbb{N}$ fixed.
If we define
    \begin{equation} \label{sin}
    \sin(\theta_0) = \frac{\|S_k|\psi^0\rangle\|}{2}
\end{equation}
and $\sin(\theta_p)$ as in \eqref{sen thetak} for $p\geq 1$, then
\begin{equation} \label{sin ineq}
	\sin(\theta_{p-1})\leq \sin(\theta_p),
\end{equation}
	for every $1\leq p\leq m$ and $m\in\mathbb{N}$.    
\end{proposition}

\begin{remark}
The formulas in \eqref{sin} and \eqref{sin ineq} are also valid for any Hermitian operator $S$ that can be written as a sum of unitary operators (see proof of Proposition \ref{eq:monotoniafiltro} in the appendix).
\end{remark}

So all subsequent probability amplitudes are lower bounded by the initial one. We can actually also bound $\sin(\theta_0)$, as we show below.

\begin{lemma}\label{lemma bound}
    	Let $|\psi^0\rangle\in \mathcal{H} $ be an initial state $\|\psi^0\|=1$ and $\sin(\theta_0$) defined in \eqref{eq:sin-theta0}. Then, the following statements hold:
    \begin{enumerate}
        \item if $\{E_i\}$ is the Hamiltonian eigenvectors, \eqref{eq:sin-theta0} takes the form
\begin{equation} \label{seno vectorial version}
    \sin(\theta_0) =\frac{1}{2}\sqrt{\sum_{i=0}^{d-1}\left(1-\sin\left(\frac{E_it}{k}\right)\right)^2|\langle\psi^0|E_i\rangle|^2}
\end{equation}
    \item It is possible to set an initial $k'\in \mathbb{N}$ such that
    \[\sin(\theta_0)=\frac{\|S_{k'}|\psi^0\rangle\|}{2}>0.\]
    \item We obtain the following lower bound for the sine function,
    \begin{equation}\label{gm>am gral}
    \sin(\theta_0)\geq \frac{1}{2\sqrt{d}}\sum_{i=0}^{d-1} \left(1-\sin\left(\frac{E_it}{k}\right)\right)|\langle\psi^0|E_i\rangle|.
\end{equation}
In particular, if we consider $|\psi^0\rangle=\frac {1}{\sqrt{d}}\sum_{j=0}^{d-1}|E_i\rangle$ as initial state, then for every $k>\frac{6\|H\|t}{\pi}$
\begin{equation}\label{gm>am partic}
        \sin(\theta_0)\geq \frac {1}{2d}\sum_{i=0}^{d-1}\left(1-\sin\left(\frac{E_it}{k}\right)\right) \geq \frac{1}{4}. 
\end{equation}

    \end{enumerate}

\end{lemma}

\section{QSITE implementations}\label{sec:implementations} 
We analyze here two possible implementations of the operator $S_k^m(t) = \left(I - \sin(tH/k)\right)^m$ that can be realized on a quantum computer. In the \emph{sequential} implementation, the operator is built by applying $m$ single-step LCU blocks one after another, each terminated by a mid-circuit measurement of two ancilla qubits that are then reset and reused. In the \emph{coherent} QSVT implementation, the operator is applied as a single unitary followed by one final measurement. As we show below, both implementations prepare the same state and succeed with identical probability. They differ in their resource requirements and, most importantly, in their compatibility with amplitude amplification. We describe each implementation in turn and then compare their computational cost. 

\subsection{Sequential QSITE}\label{sec:seq-qsite}
In the sequential implementation, the operator $S_k^m(t)$ is executed as $m$ consecutive LCU blocks. Each block encodes the single step $S_k$ using two ancilla qubits, applies it to the current system state, and measures the ancillas: the $|00\rangle$ outcome heralds a successful application of $S_{k}$, after which the ancillas are reset and reused for the next block (Sec.~\ref{sec:qsite}); any other outcome ends the shot. The register size is therefore $N+2$ independently of $m$, at the price of $m$ mid-circuit measurements and the post-selection they entail. 

At this point, it is useful to summarize the different types of probabilities we will be handling:
\begin{itemize}
   \item  Filter probability: Denoted as $\sin^2(\theta_p)$, this is the probability of successfully measuring the required ancilla state
      $|00\rangle$ at a single, specific LCU step $p$. 
   \item  Overall filter probability (also referred to as ``cumulative filter probability"): Defined as $\prod_{p=0}^{m-1}\sin^2(\theta_p)$, this
      is the combined probability of successfully passing all ancilla measurements consecutively across all $m$ concatenation steps. 
   \item Probability of failure: The probability that the final quantum state (assuming all previous LCU filters were successfully passed) yields an energy higher than the acceptable threshold ($E > E_0 + \delta E$) with $\delta E$ an acceptable tolerance. This will be better clarified in Sec.~\ref{sec:pf-qsite} when measured in the computational basis for optimization Hamiltonians. For simplicity, $P_F$ will denote the probability of failure of QSITE Eq.~(\ref{eq:pf-site-km}), unless otherwise stated. 
   \item Total probability of failure (referred to  as ``total failure probability"): Defined by the formula
	   $P_F^\mathrm{total} = 1 - \left(1 - P_F\right)\cdot\prod_{p=0}^{m-1}\sin^2(\theta_p)$. This is the practical metric: it represents the probability of not getting a correct answer in a single circuit execution (a ``shot"), taking into account both the chance that the filters fail and the chance that the measured bitstring has an energy above $E_0+\delta E$. 
\end{itemize}

When the per-step filter probability is low, amplitude amplification can boost the chance of a successful ancilla measurement. Grover-based amplitude amplification~\cite{Grover1996,Brassard2002} with $r$ iterations raises the effective success probability from $\sin^2(\theta_p)$ to $\sin^2\!\big((2r+1)\theta_p\big)$. In this work we use $r^* = \lfloor \pi/(4\theta_p) - 1/2\rfloor$, the largest $r$ for which $(2r+1)\theta_p$ does not exceed $\pi/2$, or, where stated, the smallest $r$ that raises the success probability to at least $0.5$; $r^*$ is nonzero only when $\sin(\theta_p) \le 1/2$ (equivalently, $\sin^2(\theta_p) \le 1/4$). 

One might hope to boost the overall filter probability by applying this amplification after each of the $m$ concatenation steps. This construction, however, is difficult to implement in the sequential circuit. Amplifying the filter probability at step $p$ requires reflecting about the state that enters that step, and constructing this reflection demands an explicit preparation unitary for $|\psi^{p-1}\rangle$; for $p>1$ the sequential circuit provides no such unitary, because the intermediate states are produced only by post-selection on the ancilla measurements. Within this sequential, post-selected schedule, amplitude amplification is therefore available only at the first step, whose input $|\psi^{0}\rangle = \mathbf{H}^{\otimes N}|0\rangle^{\otimes N}$ does have an explicit preparation unitary. In the numerical results of Sec.~\ref{sec:numerical-results}, the sequential schedule is always run without amplitude amplification; amplification is studied only for the QSVT implementation (Sec.~\ref{sec:qsvt-comparison}). 

Multi-step amplification instead requires abandoning the intermediate measurements altogether, so that the accumulated circuit becomes a single coherent unitary acting from $|0\rangle$: an explicit preparation unitary on which standard Grover amplitude amplification can act. One route is to render each step deterministic, as in the PITE construction combined with quantum amplitude amplification of Ref.~\cite{Nishi_2022_QAA}. This is no longer the sequential post-selected circuit, and it carries two penalties: the reflection at each step re-runs the entire accumulated history, so the cost compounds with the number of steps (growing at least as $3^{m-1}$ for $m$ steps in~\cite{Nishi_2022_QAA}), and reaching the deterministic point requires knowing the success amplitude of each step exactly, in order to tune the normalization (the parameter $\gamma$ in~\cite{Nishi_2022_QAA}). Merely deferring all ancilla measurements to the end of the sequential circuit while inserting the amplification stages between concatenations does not work either, because without the intermediate resets the next block would also act on the failed branches, introducing undesirable terms that mix ancilla and system amplitudes and spoil the filtered state. What succeeds is realizing the entire polynomial $x^m$ (with $x=S_k/2$) as a single coherent, measurement-free unitary (amplified once, from the initial state), which is precisely what the quantum singular value transformation provides, as described in the next subsection.

\subsection{QSVT implementation of QSITE}\label{sec:qsvt-impl}
The operator $S_k^m$ can instead be applied \emph{coherently}, with no intermediate measurements, through the quantum singular value transformation (QSVT). It is, by construction, a degree-$m$ polynomial in terms of $S_k$, and hence a transformation of the eigenvalues of $H$. Polynomial transformations of the eigenvalues (or singular values) of a block-encoded operator are precisely the task addressed by QSVT and the underlying quantum signal processing (QSP) framework~\cite{Low_2019_Qubitization, Gilyen_2019_QSVT, Martyn_2021_GrandUnification, Lloyd_2021_HamiltonianQSVT}, which unifies Hamiltonian simulation, amplitude amplification and matrix inversion under a single circuit template. It is therefore natural to ask whether QSVT provides a cheaper route to the QSITE state than the concatenated-LCU construction of sequential QSITE. We compare the two below, both at the level of the produced state and, more importantly, at the level of the required resources. 

The single QSITE step, Eq.~\eqref{eq:site-lcu}, is a three-term LCU with normalization $\lambda = 1+\tfrac{1}{2}+\tfrac{1}{2} = 2$. 
Encoding Eq.~\eqref{eq:site-lcu} using $\lceil\log_2 3\rceil = 2$ ancilla qubits and the usual PrepSelPrep method~\cite{Childs_2012_LCU, Berry_2015_TruncatedTaylor, Babbush_2018_LinearT}, the QSVT implementation applies a polynomial $p(x)=x^m$ with $x=S_k/2$, reproducing $(S_k/2)^m$ coherently in a single circuit. The circuit output requires only one ancilla measurement rather than the $m$ mid-circuit measurements of the sequential schedule. After real-part extraction (discussed below), the all-zero outcome of the ancillas heralds the QSITE state. In a circuit-level simulation, QSVT/PrepSelPrep and QSVT/Embedding~\cite{Gilyen_2019_QSVT}, both with real-part extraction, yield the \emph{same} normalized final state $\propto S_k^m|\psi_0\rangle$ as QSITE, and hence the same $P_F$ (Sec.~\ref{sec:qsvt-comparison}). The comparison is thus entirely about implementation cost, summarized in Table~\ref{tab:qsvt} for a representative 6-qubit instance and discussed below. 

By the telescoping identity Eq.~\eqref{eq:cumulative-filter}, the probability that all $m$ sequential ancilla measurements of QSITE succeed, $\prod_{p=0}^{m-1}\sin^2(\theta_p)$, equals $\|S_k^m|\psi^0\rangle\|^2/\lambda^{2m}$ with $\lambda = 2$, which is precisely the single-measurement success probability of the coherent QSVT circuit implementing $S^m/\lambda^m$. The sequential circuit, with $m$ measurements, and the QSVT circuit, with a single measurement after real-part extraction, are therefore \emph{exactly} equally likely to succeed---an algebraic identity rather than a numerical coincidence. Where the two implementations genuinely differ is in their compatibility with amplitude amplification: the coherent QSVT circuit is a single unitary acting on the initial state, so its success probability can be boosted by standard Grover amplitude amplification (Sec.~\ref{sec:seq-qsite}), whereas the sequential schedule admits amplification only at the first step. In regimes where the cumulative filter probability is small---in particular near the $g/2^N$ floor of Eq.~\eqref{eq:filter-floor}---this improvement makes the coherent realization the preferable route to the QSITE state; we demonstrate this numerically in Sec.~\ref{sec:qsvt-comparison}. 

\begin{table*}[t]
\centering
\begin{tabular}{l c c c c c}
\toprule
 & Sequential & \multicolumn{2}{c}{QSVT/PrepSelPrep (coherent)} & \multicolumn{2}{c}{QSVT/Embedding (coherent)} \\
\cmidrule(lr){2-2}\cmidrule(lr){3-4}\cmidrule(lr){5-6}
Resource & QSITE & raw & $+$ Hadamard & raw & $+$ Hadamard \\
\midrule
Block-encoding ancillas   & $2$           & $2$      & $2$      & $1$      & $1$ \\
Real-part extraction      & --            & --       & $1$      & --       & $1$ \\
Total qubits              & $N+2$         & $N+2$    & $N+3$    & $N+1$    & $N+2$ \\
Ancilla measurements      & $m$ (+reset)  & $1$      & $1$      & $1$      & $1$ \\
Ancilla coherence (blocks) & $1$          & $m$      & $m$      & $m$      & $m$ \\ 
Amplitude amplification   & first block only & whole circuit & whole circuit & whole circuit & whole circuit \\ 
Phase-angle synthesis     & none          & required & required & required & required \\
Imaginary leakage         & none          & yes      & removed  & yes      & removed \\
Filter probability        & $0.469$       & $0.672$  & $0.469$  & $0.672$  & $0.469$ \\ 
$P_F$ (single shot)       & $0.533$       & $0.665$  & $0.533$  & $0.665$  & $0.533$ \\
\bottomrule
\end{tabular}
\caption{Resources for the 6-qubit MaxCut instance of Fig.~\ref{fig:qsvt-vs-qsite} ($N=6$, $k=m=6$, $t=t_{E_0}$, $\delta E=1$). Sequential QSITE measures its two ancillas after each of the $m$ blocks. The two coherent QSVT realizations of $(S_k/2)^m$, with the PrepSelPrep block encoding or a generic embedding~\cite{Gilyen_2019_QSVT} of $S_k$, use a single final measurement. Measured directly (\emph{raw}), they herald a state with imaginary leakage ($P_F=0.665$); with real-part extraction (one extra ancilla) they give the QSITE state and filter probability. The embedding saves one ancilla and gains a factor $(\lambda/\alpha)^{2m}$ in success probability when $\alpha=\max_j|s_{E_j}|<\lambda=2$; here $\alpha=\lambda$, so the gain vanishes.} 
\label{tab:qsvt}
\end{table*}

 Both QSITE and QSVT/PrepSelPrep encode Eq.~\eqref{eq:site-lcu} with the same two ancilla qubits; they differ in how the $m$-th power is realized. QSITE applies $m$ blocks \emph{sequentially}, measuring the two ancillas after each block and resetting them for the next step, so the ancilla overhead is constant ($2$ qubits) regardless of $m$, at the price of $m$ mid-circuit measurements. QSVT/PrepSelPrep instead interleaves $m$ block-encoding calls with QSP phase rotations into one coherent circuit ending in a \emph{single} ancilla measurement. Trading $m$ measurements for one is what makes the whole QSVT circuit amenable to amplitude amplification; the cost is that its ancillas must stay coherent through all $m$ blocks instead of one (the system register must do so in both schemes). The embedding variant uses one fewer ancilla ($N+1$) and can in principle raise the success probability by a factor $(\lambda/\alpha)^{2m}$ when $\alpha=\max_j|s_{E_j}|<\lambda=2$ (i.e.\ for $t$ strictly inside the safe zone), but it requires a generic block encoding of $S_k$, which is far more expensive than the structured LCU of Eq.~\eqref{eq:site-lcu}. 

A more substantial distinction appears when QSVT is used to implement ITE \emph{directly}, i.e.\ to apply the genuine ITE propagator $e^{-tH}$. Since $e^{-tx}$ is not a polynomial, QSVT requires (i)~approximating it by a polynomial of degree $d$ (e.g.\ a truncated Chebyshev or Jacobi--Anger expansion), with $d$ growing with the evolution time and the target accuracy, and (ii)~synthesizing the $d+1$ QSP phase angles, a classical preprocessing step that is polynomial in $d$ and numerically delicate at high degree~\cite{Gilyen_2019_QSVT, Martyn_2021_GrandUnification}. QSITE avoids both. The sine filter in Eq.~\eqref{eq:site-lcu} is \emph{not} an approximation of a transcendental function: it arises exactly from the LCU of the forward and backward short-time evolutions $e^{\pm itH/k}$, and the only ``angles'' involved are the analytic times $\pm t/k$. The controlled QSITE--ITE approximation is governed instead by the parameters $k$ and $m$ through Theorems~\ref{teo:site-operator} and~\ref{teo:site-state}, with no polynomial fitting and no angle-finding subroutine. For a diagonal MaxCut Hamiltonian the evolutions $e^{\pm itH/k}$ are moreover \emph{exact} products of $ZZ$-rotations (Section~\ref{sec:complexity}), so the QSITE circuit introduces no Trotter or truncation error. 

Measured directly, the QSVT circuit does not give $x^m$ alone: with a generic choice of phases its block is a complex polynomial $x^m+i\,q(x)$, and for even $m$ the imaginary part $q$ cannot vanish~\cite{Gilyen_2019_QSVT}. This imaginary leakage degrades the result: in the 6-qubit example the single-shot $P_F$ rises from $0.53$ to $0.67$ (Fig.~\ref{fig:qsvt-vs-qsite}, right). The real part is recovered exactly with one additional ancilla (a Hadamard-test construction~\cite{Gilyen_2019_QSVT, Martyn_2021_GrandUnification}) and no extra block-encoding queries; the filter probability is then the QSITE value ($0.469$ instead of $0.672$). The QSITE step of Eq.~\eqref{eq:site-lcu} is real by construction, so it has no leakage. 

In summary, sequential QSITE and the QSVT realizations with real-part extraction prepare the same state with the same filter probability. Sequential QSITE uses a fixed two-ancilla circuit with no phase-angle synthesis and no imaginary leakage, but needs $m$ mid-circuit measurements and allows amplitude amplification of the first block only. The QSVT realizations need a single measurement, phase angles and one extra ancilla for real-part extraction, and allow amplitude amplification of the whole circuit, the property exploited in Sec.~\ref{sec:qsvt-comparison}.

\section{Optimization Hamiltonians and Probability of Failure}\label{sec:optimization-based-on-qite}
In \cite{PhysRevA.111.042403} a method using QITE was proposed for obtaining ground states of optimization hamiltonians of the form

\begin{equation}
\label{eq:H_optimization}
H = a I +  \sum_{i= 0}^{N-1} b_i Z_i + \sum_{(i, i')}^{N-1} c_{i,i'} Z_i Z_{i'}, 
\end{equation}
with $I$ the identity operator, $Z_i$ the $Z$ Pauli matrix acting on the i-th qubit, and $a$, $b_i$, $c_{i,i'}$ real coefficients.

 The problem to be solved was cast in terms of tolerable error $\delta E$. That is, to find the lowest eigenvalue $E_0$ of $H$ or an eigenvalue $E_i$ such
that $E_i \leq E_0 + \delta E$, with $\delta E$ a tolerable error. One starts with an initial state of the form $|\psi_0\rangle=\mathbf{H}^{\otimes N}|0\cdots0\rangle$ (with $\mathbf{H}^{\otimes N}$ the Hadamard gate acting on each qubit).
Apply the QITE operator of \cite{Motta_2019} on the initial state up to maximum number of iterations and then measure $S$ times (number of shots) in the computational basis. Then choose from the $S$ output the state with less energy. If the energy associated to the state with less energy satisfies to be lower than $E_0 + \delta E$ then this state is an acceptable state to solve the optimization problem.

Since the proposed method is probabilistic, there is a  failure probability associated with it, and the lower it is, the higher the chances of obtaining at least one acceptable eigenvalue of $H$ after $S$ shots.
Given the QITE state at time $t$,  $\ket{\phi^{qite}_t}$, when measuring one time ($S=1$) in the computational basis, the \textit{failure probability} is the probability of obtaining an eigenvalue $E_i > E_0+\delta E$, which is given by 
\begin{equation}\label{eq:qite-failure-probality}
        P_F^{qite}(t)= \sum_{E_i > E_0+\delta E }
        | \langle E_i \ket{\phi^{qite}_t} | ^2.   
\end{equation}
When measuring $S$ times in the computational basis, the failure probability for the QITE state at time $t$ is $(P_F^{qite}(t))^S$. Similarly, it can be defined for the ITE and QSITE states. 

\subsection{Probability of Failure of QSITE}\label{sec:pf-qsite}

The operator $S_k^m(t)$ allows us to better understand the results given by concatenation $m$ of $U_{LCU}$ circuits after every passed filter. 
  Let us consider the ordered set of eigenvalues of H that are greater than $E_0+\delta E$, for some $\delta E>0$. That is, $\{E_i: E_i > E_0 + \delta E\}$ for $i:r,\cdots,d-1$. Here $r$ is some index between $0,\cdots, d-1$.  We define the  failure probability for QSITE as: 

    \begin{equation}
	    P^{qsite}_{F,~k,m}(t)=\sum^{d-1}_{i=r}|\langle E_i|\psi^{qsite}_{k,m}(t)|^2. 
    \end{equation}

    Let the initial state be $|\psi_0\rangle=\frac{1}{\sqrt{d}}\sum^{d-1}_{i=0}|E_i\rangle$ with $d=2^N$. We define $s^{m}_{E_{i},k}(t)$ as
    \begin{equation}
	s^{m}_{E_{i},k}(t)=\left(1-\sin\left(\frac{E_it}{k}\right)\right)^m.
    \end{equation}
 Then, the QSITE state at time $t$ with parameters $k$ and $m$ is defined as: 
    \begin{align}
	    |\psi^{qsite}_{k,m}(t)\rangle &=\frac{S_k^m(t)|\psi_0\rangle}{||S_k^m(t)|\psi_0\rangle||}=\frac{\sum^{d-1}_{i=0}s^m_{E_{i},k}(t)|E_i\rangle}{\sqrt{\sum^{d-1}_{i=0}s^{2m}_{E_{i},k}(t)}}. 
    \end{align}
We see that:
    \begin{equation}\label{eq:pf-site-km}
	    P^{qsite}_{F,~k,m}(t)=\frac{\sum^{d-1}_{i=r}s^{2m}_{E_{i},k}(t)}{\sum^{d-1}_{i=0}s^{2m}_{E_{i},k}(t)}=\frac{\sum^{d-1}_{i=r}s^{2m}_{E_{i},k}(t)}{\sum^{r-1}_{i=0}s^{2m}_{E_{i},k}(t)+\sum^{d-1}_{i=r}s^{2m}_{E_{i},k}(t)}. 
    \end{equation}

\subsection{MAXCUT}
Let $G = (V,E)$ be a graph with vertex set $V$ and edge set $E$, and let $N$ be the number of vertices of the graph $G$. A maxcut hamiltonian associated to a graph $G$ gets the following form:

\begin{equation}\label{eq:maxcut-ham}
    H=-\sum_{(i,j)\in E}\frac{I-Z_iZ_j}{2},
\end{equation}

all the eigenvalues are non-positive integers. 

\begin{proposition}\label{eq:pfsitemaxcutlinearinN}
    Let $\{E_i\}$ be the ordered set of eigenvalues of a maxcut hamiltonian and $t_{E_0}=\frac{\pi}{2}\frac{k}{|E_0|}$, with $E_0$ the lowest eigenvalue. We then have 
    \begin{equation}
    P^{qsite}_{F,~k,m}(t_{E_0})\leq \frac{(d-r)s^{2m}_{E_r,k}(t_{E_0})}{g~2^{2m}+(d-r)s^{2m}_{E_r,k}(t_{E_0})}, 
    \end{equation}
with $s_{E_r,k}(t_{E_0})<2$ and $g$ the degeneracy of $E_0$. More generally, for every $t\in(0,t_{E_0}]$ the same bound holds with $t_{E_0}$ replaced by $t$ and $2$ by $s_{E_0,k}(t)$, where $s_{E_r,k}(t)<s_{E_0,k}(t)$. 
\end{proposition}
Proposition~\ref{eq:pfsitemaxcutlinearinN} implies that around $t_{E_0}=\frac{\pi}{2}\frac{k}{|E_0|}$ the probability of failure of QSITE can be decreased if we increase $m$. It is easy to see that if we take $m$ to satisfy: 

\begin{equation}
    2m\left(1-\log_2(s_{E_r,k}(t_{E_0}))\right) + \log_2(g) - N \geq 0, 
\end{equation}
and since $s_{E_r,k}(t_{E_0})<2$, such an $m$ always exists and we obtain an upper bound of $1/2$ for $P^{qsite}_{F,~k,m}$. If, in addition, $\delta E$ is a fixed fraction of $|E_0|$, i.e.\ $\delta E=f|E_0|$ with $0<f<1$, then $|E_r|<(1-f)|E_0|$ and therefore $s_{E_r,k}(t_{E_0})<1+\cos(\pi f/2)<2$ uniformly in $N$, so we see that $m=O(N)$. That is, it is enough to let $m$ to be increased linearly with the number of qubits to obtain a fixed probability of failure. 

The form of $P^{qsite}_{F,~k,m}(t)$ is quite complicated and is periodic in $2\pi k$. In fact, there exist many points such that the probability of failure can be made very low. Unfortunately, we don't have a controlled way to obtain those values except for small intervals around $t_{E_0}$.

\subsection{Computational complexity}\label{sec:complexity}

We now analyze the resources required by sequential QSITE for a MaxCut Hamiltonian.

The circuit acts on $N$ system qubits and two ancillas. The ancillas are measured after each block and, on the outcome $|00\rangle$, reset and reused, so the register size is $N+2$ for any $m$.

The only non-trivial operations inside each LCU block are the controlled evolutions $e^{\pm i tH/k}$ and the two-qubit Householder gate $V$. For a MaxCut Hamiltonian, Eq.~\eqref{eq:maxcut-ham}, $H$ is diagonal and can be written as $H=-\frac{|E|}{2}I+\frac{1}{2}\sum_{(i,j)\in E}Z_iZ_j$, so
\begin{equation}
    e^{-i\frac{t}{k}H}=e^{i\frac{t|E|}{2k}}\prod_{(i,j)\in E}e^{-i\frac{t}{2k}Z_iZ_j},
    \label{eq:maxcut-evolution}
\end{equation}
a product of $|E|$ commuting $ZZ$-rotations, one per edge of $G=(V,E)$, and a global phase. Each rotation takes two CNOTs and one single-qubit rotation, and the control on the ancillas adds a constant overhead per gate, so each LCU block uses $O(|E|)$ two-qubit gates. With $m=O(N)$ steps, which give $P_F\le 1/2$ at $t=t_{E_0}$ when $\delta E$ is a fixed fraction of $|E_0|$ (Proposition~\ref{eq:pfsitemaxcutlinearinN}), the full circuit uses $O(N|E|)$ two-qubit gates: $O(N^3)$ for dense Erd\"os--R\'enyi graphs $G(N,1/2)$, where $|E|=\Theta(N^2)$, and $O(N^2)$ for graphs of bounded degree.

A single execution succeeds only when all $m$ ancilla measurements yield $|00\rangle$ \emph{and} the measured bitstring has energy at most $E_0+\delta E$. The success probability of one shot is $p_{\mathrm{succ}}=(1-P_F)\prod_{p=0}^{m-1}\sin^2(\theta_p)$, so the expected number of repetitions is $p_{\mathrm{succ}}^{-1}$. With $m=O(N)$ as above, $1-P_F\ge 1/2$, and the sampling cost is set by the cumulative filter probability. Squaring Eq.~\eqref{norma S^k} gives
\begin{equation}\label{eq:cumulative-filter}
\prod_{p=0}^{m-1}\sin^2(\theta_p)=\frac{\left\|S_k^m|\psi^0\rangle\right\|^2}{2^{2m}},
\end{equation}
for every $m$. At $t=t_{E_0}$ the ground-state coefficient reaches the maximum value $s_{E_0,k}(t_{E_0})=1-\sin(-\pi/2)=2$, and no other level reaches it, so $(S_k/2)^m|\psi^0\rangle$ converges to the ground-state component of the initial state, and the cumulative filter probability decreases monotonically to
\begin{equation}\label{eq:filter-floor}
\lim_{m\to\infty}\prod_{p=0}^{m-1}\sin^2(\theta_p)=\left\|\Pi_{E_0}|\psi^0\rangle\right\|^2=\frac{g}{2^N},
\end{equation}
where $\Pi_{E_0}$ is the projector onto the ground-state subspace and $g$ its degeneracy; the last equality holds for $|\psi^0\rangle=|+\rangle^{\otimes N}$ (proof in Appendix~\ref{app:filter-floor-proof}). Hence $p_{\mathrm{succ}}\ge(g/2^N)(1-P_F)$ for every $m$, and the expected number of repetitions approaches $2^N/g$ as $m$ grows, which is exponential in $N$ when $g$ is bounded. This is the regime where amplitude amplification becomes relevant (Secs.~\ref{sec:beyond-tsafe} and~\ref{sec:qsvt-comparison}). For the fully connected graphs of Appendix~\ref{app:fc}, instead, $g/2^N=\binom{N}{N/2}/2^N=\Theta(1/\sqrt{N})$, which explains their low shot counts.

\textit{Choice of the evolution time.} The time $t$ is the main hyperparameter of QSITE: not every value of $t$ yields a controlled failure probability, so $t$ must lie in one of the two safe zones, the first up to $t_{E_0}=\frac{\pi}{2}\frac{k}{|E_0|}$ (Proposition~\ref{eq:pfsitemaxcutlinearinN}) and the second around $t_2=\frac{5\pi}{2}\frac{k}{|E_0|}$ (Sec.~\ref{sec:beyond-tsafe}). Both depend on $|E_0|$, the maximum cut of the graph, which is in general unknown. Since $|E|/2\le|E_0|\le|E|$ (a random cut contains half of the edges on average), the time $t=\frac{\pi}{2}\frac{k}{|E|}$ satisfies $t_{E_0}/2\le t\le t_{E_0}$ and is obtained from the graph data without diagonalizing $H$. For a connected graph, the Edwards bound $|E_0|\ge|E|/2+(N-1)/4$~\cite{Edwards_1973,Edwards_1975} sharpens this to $\frac{\pi}{2}\frac{k}{|E|}\le t_{E_0}\le\frac{\pi}{2}\frac{k}{|E|/2+(N-1)/4}$, which locates $t_{E_0}$ from the graph data alone.

\textit{Comparison with QITE.} The original QITE algorithm of Motta \emph{et al.}~\cite{Motta_2019} is ancilla-free but reproduces each non-unitary step by a unitary acting on a domain of $D$ qubits, whose size grows with the correlation length $C$ of the evolving state ($D=O(C^d)$ on a $d$-dimensional lattice). This unitary must be reconstructed at every step from measured expectation values and a least-squares solve, and the measurements, the solve and the gate synthesis each cost $\exp(O(C^d))$~\cite{Motta_2019}, which is efficient for short-range correlations. For the dense, non-geometric MaxCut instances considered here the domain can extend over the whole system, $D\sim N$, making this cost exponential in $N$; QITE has nevertheless been applied to MaxCut~\cite{alam2023solvingmaxcutquantumimaginary,PhysRevA.109.052430} and UDMIS~\cite{PhysRevA.111.042403} as a heuristic with constrained domain size. QSITE needs no reconstruction: each step is a fixed circuit of $O(|E|)$ gates built from Eq.~\eqref{eq:maxcut-evolution}, with no tomography and no classical optimization, at the price of two ancillas and the sampling overhead discussed above.

QSITE builds each step from the real-time evolutions $e^{\pm itH/k}$, with a fixed count of two ancillas. Block-encoding $H$ directly, as in the Euler filter $(I-tH/k)^m$, gives a monotone filter, but its LCU needs an ancilla register growing as $\lceil\log_2(|E|+1)\rceil$. A filter built from $e^{\pm itH/k}$ is instead $2\pi k$-periodic in $t$ for MaxCut and cannot be monotone; the resulting oscillation of $P_F(t)$ (Figs.~\ref{fig:6q-Pfsite} and~\ref{fig:20q-Pfsite}) is the price of the fixed ancilla count, and it is what makes the safe-zone analysis of Sec.~\ref{sec:numerical-results} necessary.

\section{Numerical results}\label{sec:numerical-results}

We now present numerical results for QSITE applied to random MaxCut Hamiltonians of the form given in Eq.~(\ref{eq:maxcut-ham}). These hamiltonians possess a non-positive energy spectrum. For each instance, the graph $G=(V,E)$ is generated by independently including each possible edge with probability $1/2$. All simulations start from the equal superposition state $|\psi_0\rangle = \frac{1}{\sqrt{d}}\sum_{i=0}^{d-1}|E_i\rangle$ with $d=2^N$.  We present results for $N=6$, $N=12$, and $N=20$ qubits as representative cases; all other qubit counts from $N=6$ to $N=24$ display the same qualitative behavior. We analyze several complementary aspects of the algorithm: (i) the probability of failure $P^{qsite}_{F,\,k,m}(t)$ as a function of  time, (ii) the filter probability $\sin^2(\theta_p)$ and $P_F$ as functions of the number of concatenation steps $m$, (iii) the required number of measurement shots as a function of system size $N$, and (iv) the exploitation of additional safe zones beyond $t_{E_0}$. Results for fully connected MaxCut graphs are presented in Appendix~\ref{app:fc}. 

\subsection{Probability of Failure as a Function of Time}

\begin{figure}[H]
    \centering
    \includegraphics[width=0.55\linewidth]{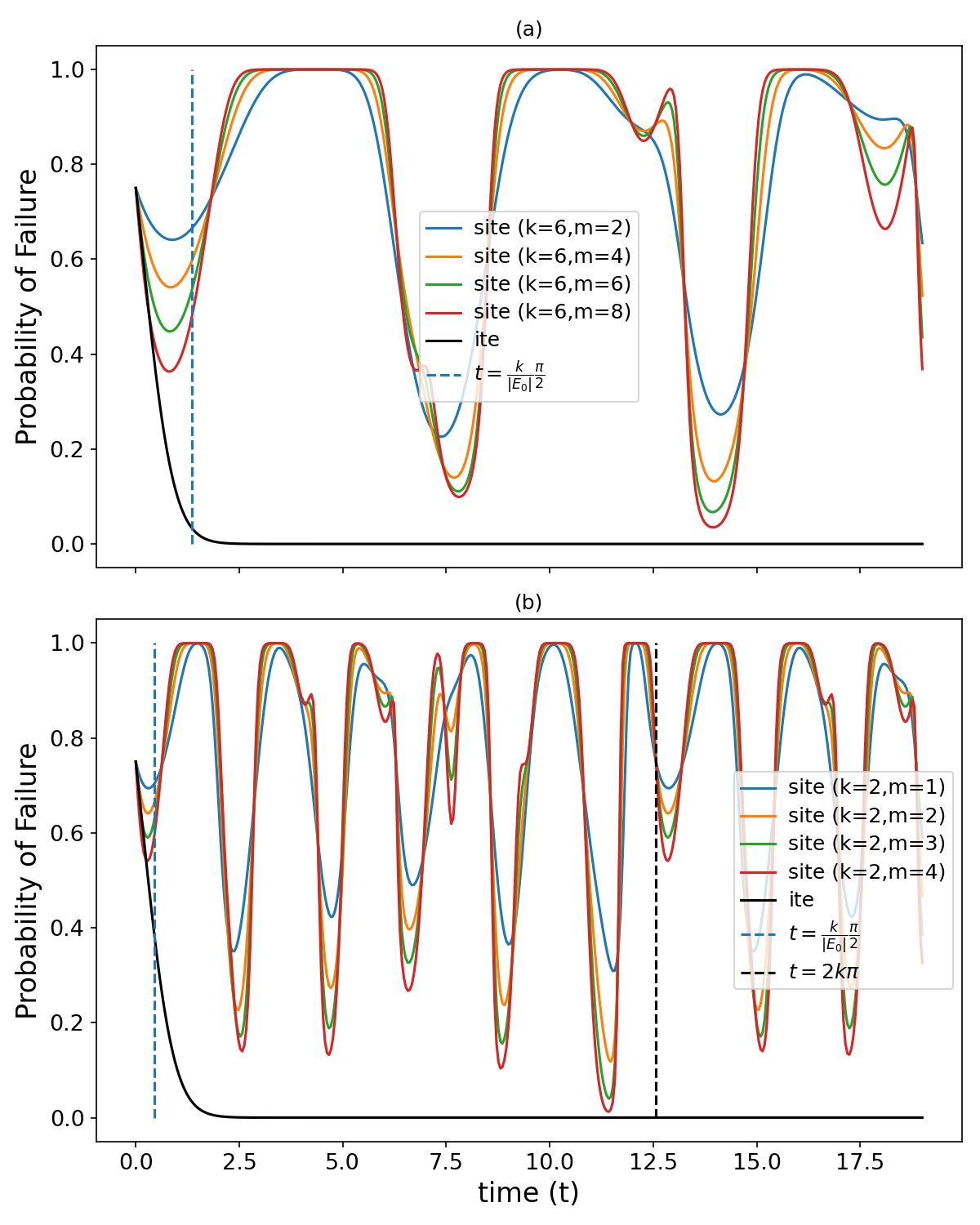}
    \caption{Probability of failure $P^{qsite}_{F,\,k,m}(t)$ as a function of time for a 6-qubit random MaxCut instance ($|E|=9$ edges, $E_0=-7$, $\delta E \approx 10\%\,|E_0|$). (a)~$k=6$ with $m=2,4,6,8$ concatenations. (b)~$k=2$ with $m=1,2,3,4$. The black curve shows the ITE reference $P^{ite}_F(t)$, and the dashed vertical line marks the safe-zone boundary $t_{E_0}=\frac{\pi}{2}\frac{k}{|E_0|}$. Increasing $m$ reduces $P_F$ in the first safe-zone, and the periodic structure with period $2\pi k$ (arising from the integer eigenvalues of MaxCut Hamiltonians) is clearly visible. The natural maximum concatenation for each case is $m=k$ (the regime in which the main convergence theorems apply); nevertheless, going beyond this value -- as illustrated with $m > 6$ for $k=6$ and $m > 2$ for $k=2$ -- further reduces $P_F$ without theoretical obstruction.} 
    \label{fig:6q-Pfsite}
\end{figure}

\begin{figure}[H]
    \centering
    \includegraphics[width=0.64\linewidth]{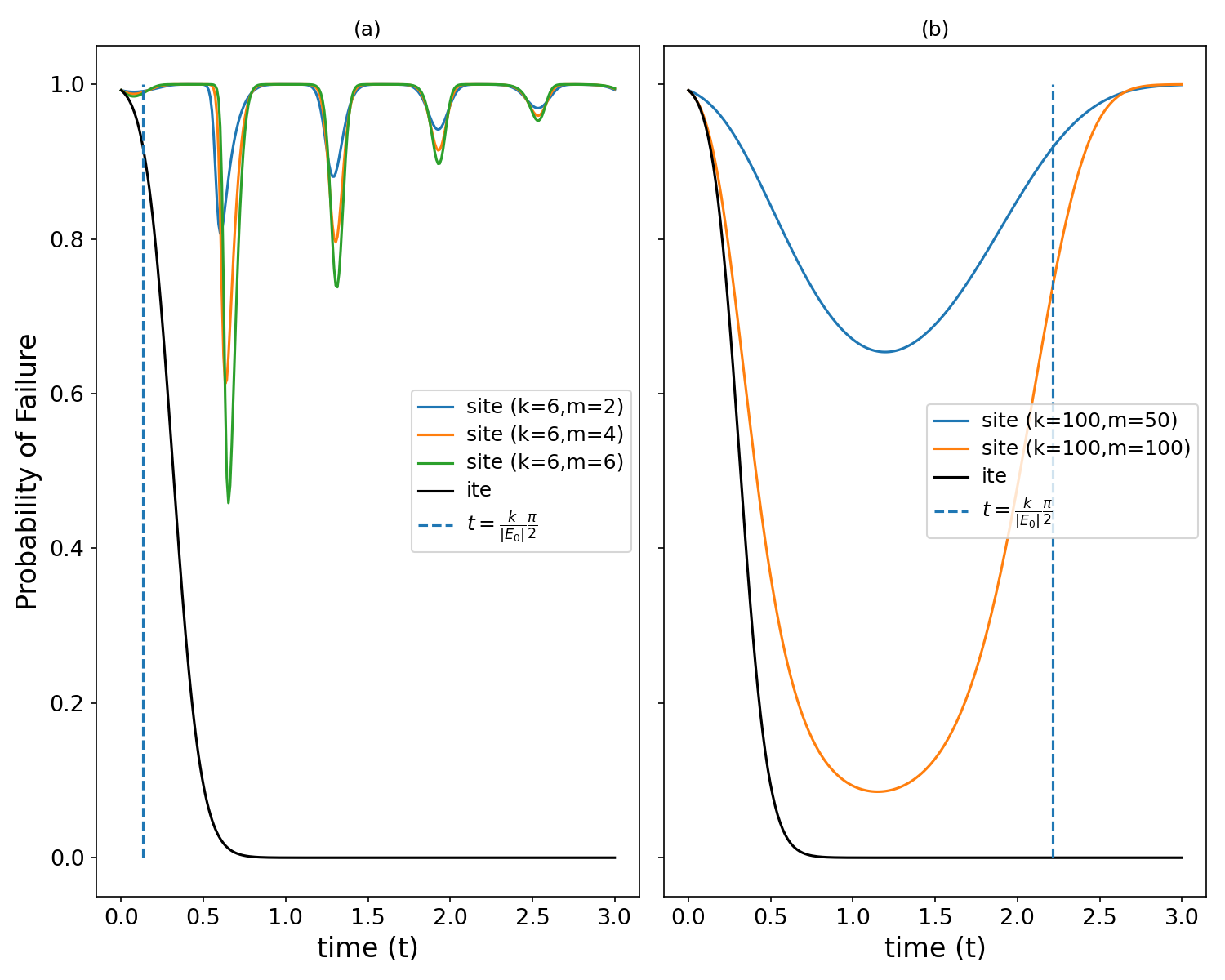}
    \caption{Probability of failure for a 20-qubit random MaxCut instance ($|E|=105$ edges, $E_0=-71$, $\delta E \approx 10\%\,|E_0|$). (a)~$k=6$, $m=2,4,6$. (b)~$k=100$, $m=50, 100$. At larger system sizes the spectral density increases and more concatenations are needed to suppress $P_F$. With $k=100$ and $m=100$ the failure probability reaches values below $0.1$ before $t_{E_0}$. When $k=m$ and both are large, the QSITE curves begin to approach the ITE reference (black), in agreement with Theorems~\ref{teo:site-operator} and~\ref{teo:site-state}.} 
    \label{fig:20q-Pfsite}
\end{figure}

Figures~\ref{fig:6q-Pfsite} and~\ref{fig:20q-Pfsite} display $P^{qsite}_{F,\,k,m}(t)$ for a 6-qubit  and a 20-qubit  random MaxCut instance, respectively, alongside the ITE reference $P^{ite}_F(t)$. The probability of failure exhibits a periodic structure with period $2\pi k$, in agreement with the periodicity of the filters  $s^m_{E_i,k}(t) = (1-\sin(E_i t / k))^m$. Within each period, the first minimum in $P_F$ occurs before $t_{E_0} = \frac{\pi}{2}\frac{k}{|E_0|}$ (blue dashed vertical line), where the ground-state component is maximized ($s_{E_0,k}(t_{E_0}) = 1 - \sin(-\pi/2) = 2$). We call the interval $[0,t_{E_0}]$ the first safe-zone. Which represents a controllable zone to study $P_F$. It can be seen from these figures that increasing the number of concatenations $m$ systematically reduces $P_F$ at these optimal time values, consistent with the bound of Proposition~\ref{eq:pfsitemaxcutlinearinN}. The probability of failure of ITE (black curve) decays monotonically to zero as expected. Thus, QSITE captures the essential behavior of ITE, albeit with oscillatory corrections arising from the sine approximation. As the values of $k$ and $m$ increase, the  QSITE curve fits the ITE curve more closely (Fig.~\ref{fig:20q-Pfsite}b) as expected from Theorem~\ref{teo:site-operator}. 

A comparison of different $k$ values reveals that larger $k$ extends the useful time range (since $t_{E_0} \propto k$) but requires correspondingly more concatenations to achieve a given $P_F$, reflecting the trade-off between the approximation quality per step and the number of steps required.

\subsection{Filter Probability and Dependence on Concatenation Steps} 

\begin{figure}[H]
    \centering
    \includegraphics[width=0.8\linewidth]{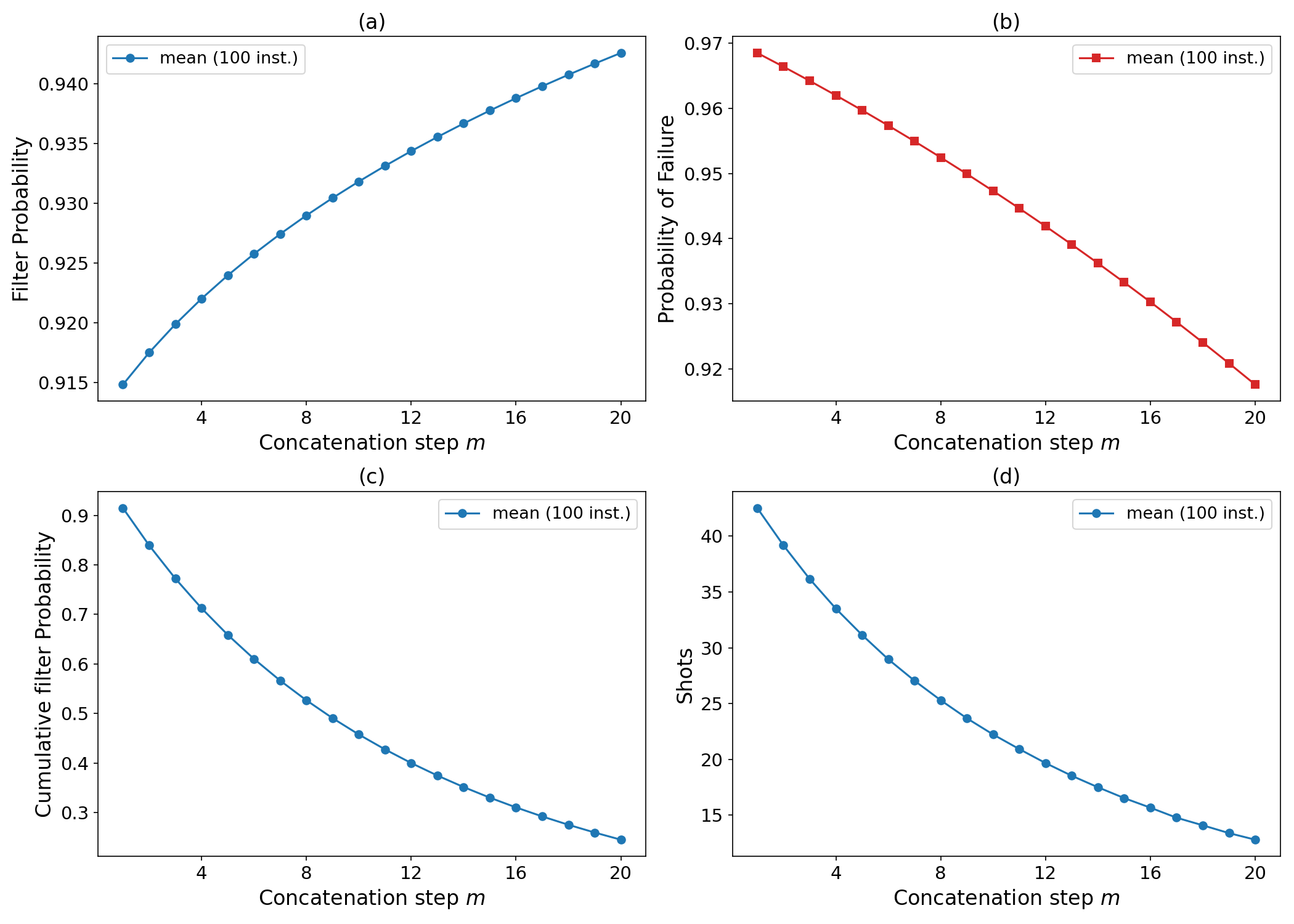}
	  \caption{Filter probability of the $m$-th step (a), failure probability $P_F$ (b), cumulative filter probability, Eq.~\eqref{eq:cumulative-filter} (c), and number of shots $S$ with $(P_F)^S<0.5$, ignoring filter failures (d), versus the number of steps $m$, for 100 20-qubit random MaxCut instances at $t=t_{E_0}$ ($k=20$, $\delta E=8\approx12\%\,|E_0|$).}
    \label{fig:filter-prob-vs-concat-20q-k20}
\end{figure}

Figure~\ref{fig:filter-prob-vs-concat-20q-k20} shows how the probabilities of Sec.~\ref{sec:seq-qsite} depend on the number of steps $m$ at $t=t_{E_0}$, for 100 20-qubit instances.
The filter probability increases with $m$ (a), as guaranteed by Proposition~\ref{eq:monotoniafiltro}, and $P_F$ decreases monotonically (b), although its mean is still above $0.9$ at $m=20$. 
Although each measurement becomes more likely to succeed, the cumulative filter probability decreases (c), from $0.91$ at $m=1$ to $0.25$ at $m=20$, and this product, not the per-step probability, sets the sampling cost. Panel (d) gives the number of shots $S$ with $(P_F)^S<0.5$, ignoring filter failures.

We have checked that the observations made do not depend on the number of qubits: the same qualitative trends seen for the 20-qubit ensemble in Fig.~\ref{fig:filter-prob-vs-concat-20q-k20} also hold for the rest of the qubits ensembles. As long as $t$ is evaluated within the safe zone $t \in [0, t_{E_0}]$ the observations hold. Also, at $t=t_{E_0}$ the coefficients $s_{E_i,k}(t_{E_0})=1+\sin\!\big(\tfrac{\pi}{2}|E_i|/|E_0|\big)$ do not depend on $k$, so these results hold for any $k$.

A more complete resource measure is the total failure probability $P_F^\mathrm{total}$, which combines the ancilla measurement success and the quality of the post-selected state. Figure~\ref{fig:joint-20q} in Appendix~\ref{app:joint-20q} shows this quantity and the corresponding number of shots for the same instances. Both quantities increase approximately linearly with $m$: the improving $P_F$ is nearly cancelled, and slightly outweighed, by the decreasing cumulative filter probability. Within the first safe zone, each additional concatenation step therefore buys a better post-selected state only at a higher sampling cost, so the overall resource budget does not improve. As we show in Sec.~\ref{sec:qsvt-comparison}, this changes once amplitude amplification is incorporated through the coherent QSVT implementation, where the filter cost can be pinned so that $P_F^\mathrm{total}$ follows the decreasing $P_F$ instead.

\subsection{Beyond $t_{E_0}$: Exploiting the Periodic Structure}\label{sec:beyond-tsafe}

The probability of failure $P^{qsite}_{F,\,k,m}(t)$ is a complex periodic function of  time $t$, as visible in Figs.~\ref{fig:6q-Pfsite} and~\ref{fig:20q-Pfsite}. While the first safe zone at $t_{E_0}$ provides a theoretically controlled regime, the periodic structure of $P_F(t)$ implies the existence of additional time values where the failure probability can be made even smaller. 

\begin{figure}[H]
    \centering
    \includegraphics[width=0.72\linewidth]{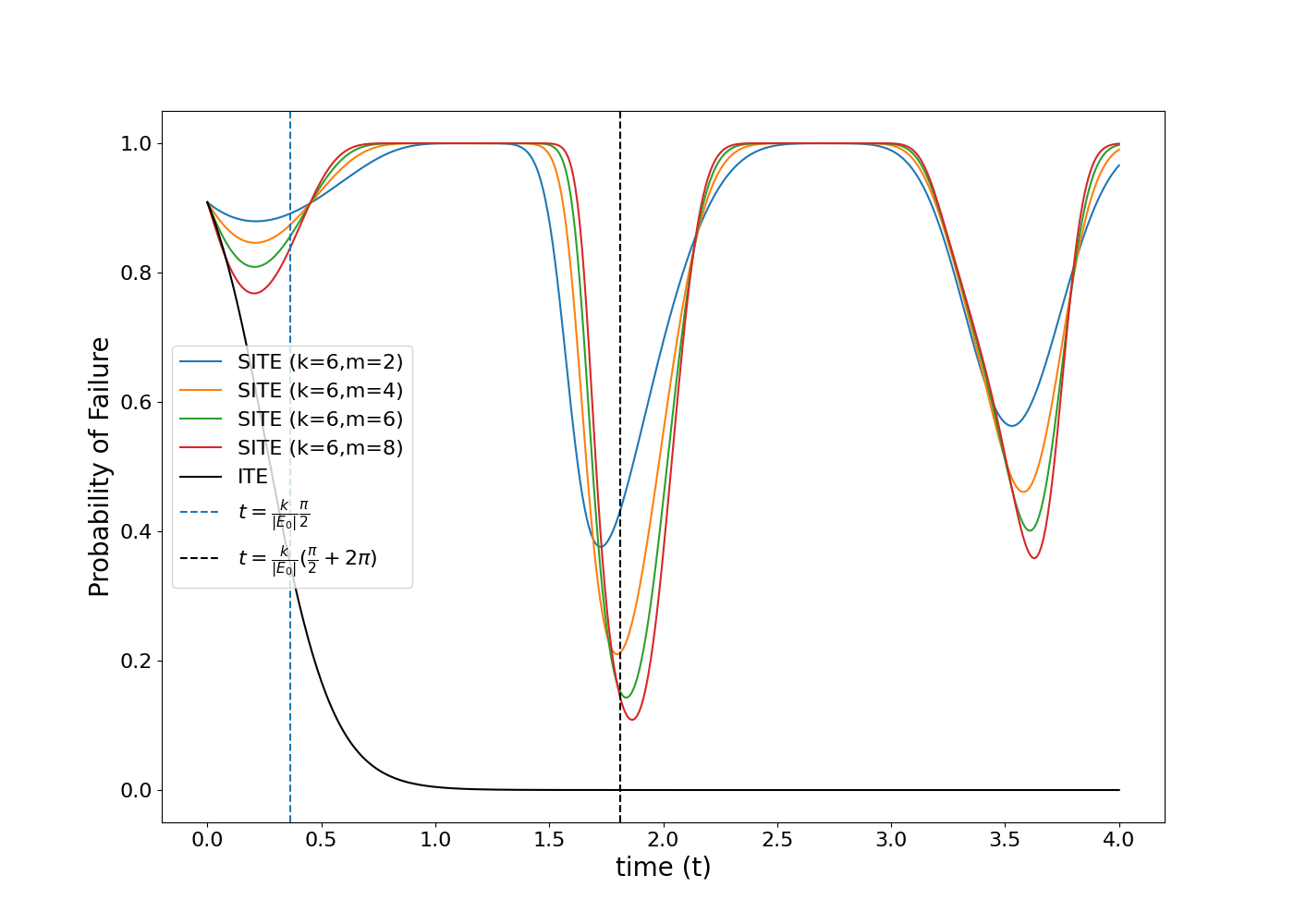}
    \caption{Probability of failure $P^{qsite}_{F,\,k,m}(t)$ vs.\ time $t$ for a 12-qubit MaxCut instance (seed 0, $\delta E = 3 \approx 10\%\,|E_0|$, $k=6$, $m=2,4,6,8$). The second safe zone provides lower $P_F$ than the first zone for the same number of concatenations.} 
    \label{fig:pf-pipeline-12q}
\end{figure}

Figure~\ref{fig:pf-pipeline-12q} displays $P^{qsite}_{F,\,k,m}(t)$ as a function of $t$ for 12-qubit instances respectively. In addition to the first minimum near $t_{E_0}$, a second deeper minimum appears at roughly $t_2 \approx \frac{5\pi}{2}\frac{k}{|E_0|}$. This second safe zone exists consistently across all qubit sizes examined, suggesting it is a robust feature of the QSITE filter structure for MaxCut instances with integer-valued eigenvalues. 

\begin{figure}[H]
    \centering
    \includegraphics[width=0.75\linewidth]{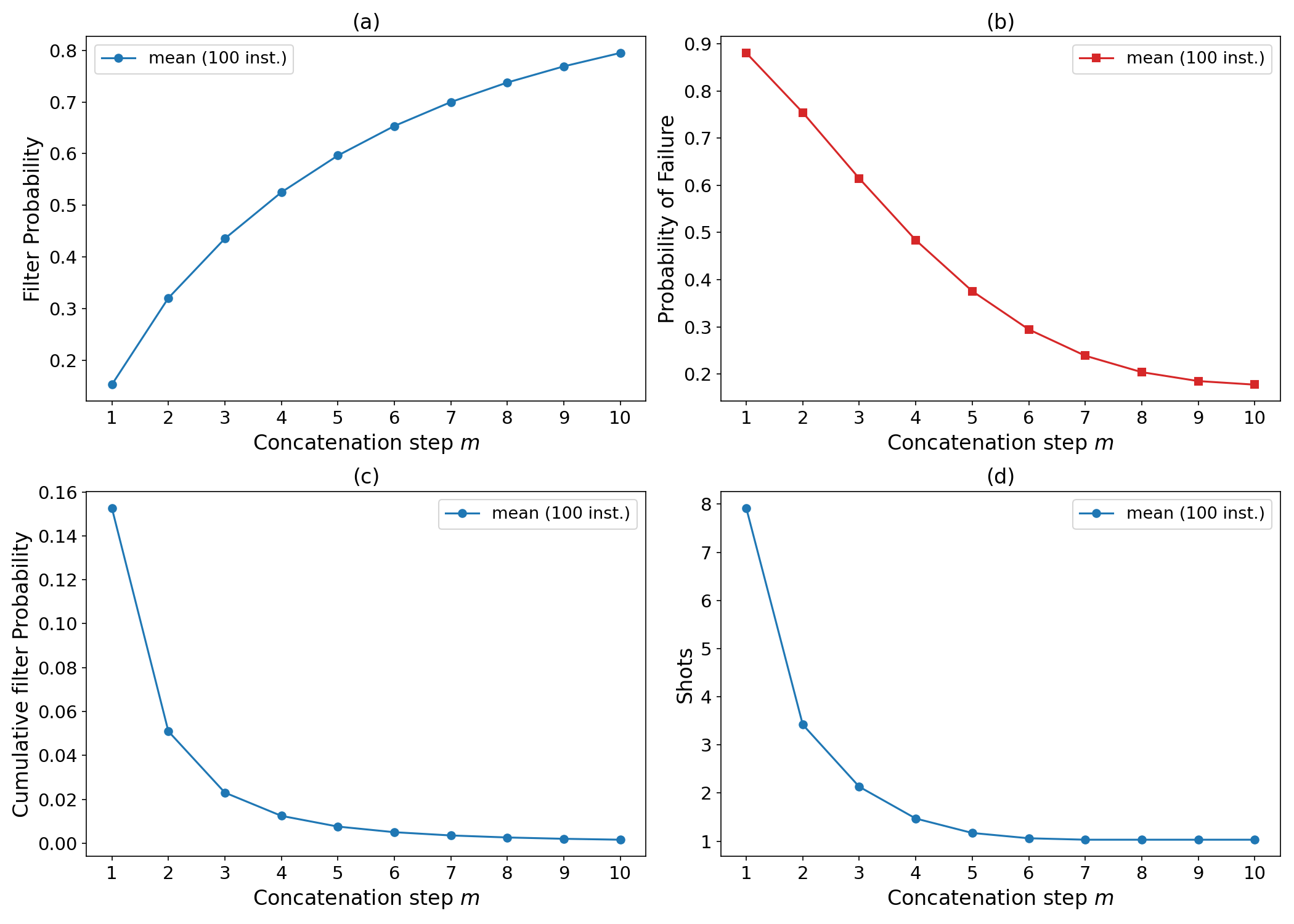}
    \caption{Filter probability (top left), failure probability $P_F$ (top right), cumulative filter probability (bottom left), and shots (bottom right) as a function of concatenation step $m$, for the 20-qubit MaxCut ensemble (100 instances, $\delta E=8$, $k=20$) evaluated at the second safe zone. Compared with Fig.~\ref{fig:filter-prob-vs-concat-20q-k20}, $P_F$ reaches lower values but the cumulative filter probability is reduced, revealing a tradeoff between state quality and circuit success rate.}
    \label{fig:grid-zone2-20q}
\end{figure}

Returning to the 20-qubit case study in the previous subsection, Figure~\ref{fig:grid-zone2-20q} presents the four-panel analysis  evaluated at the second safe zone. Compared with the first-zone results (Fig.~\ref{fig:filter-prob-vs-concat-20q-k20}), there is a notable tradeoff: while $P_F$ reaches lower values at the second zone for the same number of concatenations, the cumulative filter probability (c) is substantially reduced, implying that more ancilla measurements fail and the effective probability of successfully running the circuit decreases. This tradeoff between improved state quality and lower circuit success probability is a fundamental feature of the QSITE algorithm that must be considered when choosing the operating time $t$. 

In both safe zones, the lower $P_F$ reached with more concatenations is paid for by a lower cumulative filter probability, the probability of passing all $m$ ancilla measurements; the second zone only makes this cost larger (Fig.~\ref{fig:grid-zone2-20q}). Amplitude amplification can offset this cost, but the sequential schedule admits it only at its first step, because the post-selected intermediate states have no preparation unitary (Sec.~\ref{sec:seq-qsite}). The coherent QSVT realization of $S^m$ (Sec.~\ref{sec:qsvt-impl}), by contrast, is a single unitary acting on the initial state and admits standard Grover amplitude amplification (Sec.~\ref{sec:qsvt-comparison}).

\subsection{Amplitude amplification with the QSVT implementation}\label{sec:qsvt-comparison}
Sections~\ref{sec:seq-qsite} and~\ref{sec:qsvt-impl} established that the sequential and coherent (QSVT, with real-part extraction) implementations of $S_k^m$ prepare the same state with the same filter probability, differing only in their structural resources and in their compatibility with amplitude amplification. We now illustrate both facts numerically: first the state-level equivalence, then the amplitude-amplification payoff that only the coherent circuit admits. 

\begin{figure}[H]
    \centering
    \includegraphics[width=0.95\linewidth]{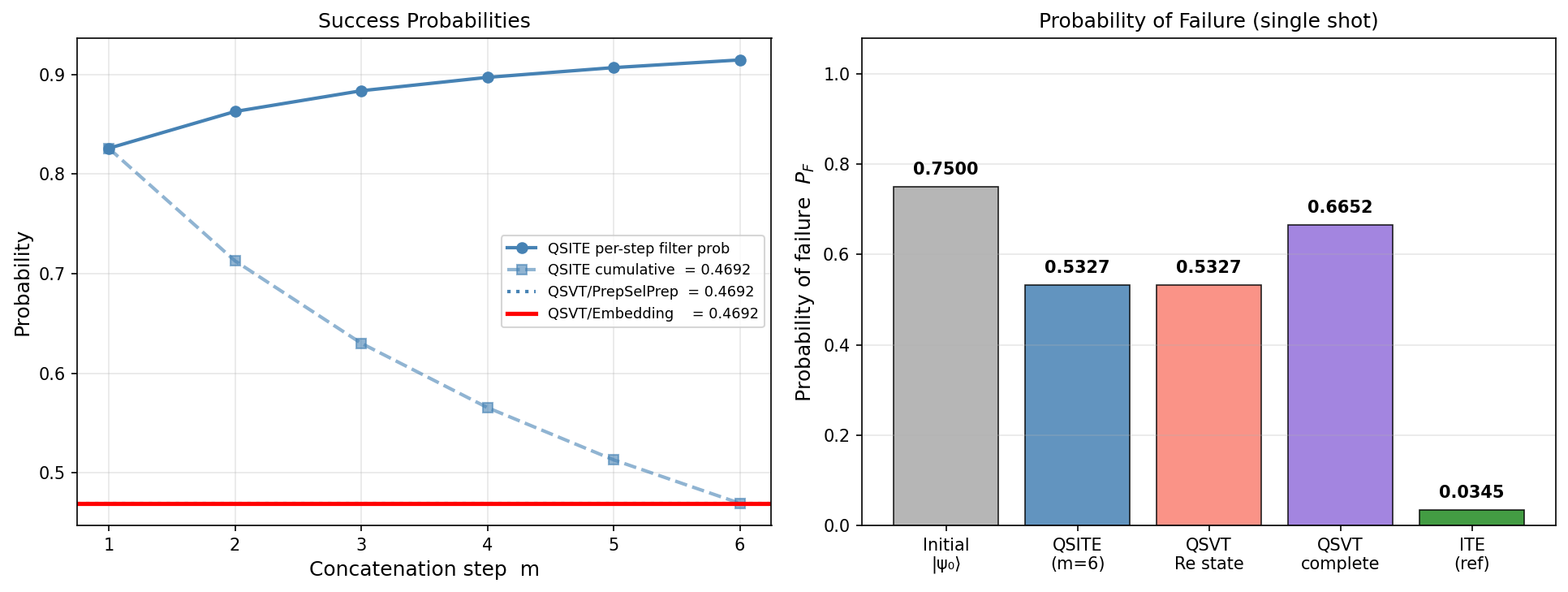} 
    \caption{QSITE versus QSVT for a 6-qubit random MaxCut instance ($k=m=6$, $t$ at the first safe-zone boundary $t_{E_0}$, $\delta E = 1$). \emph{Left:} per-step filter probability $\sin^2(\theta_p)$ and cumulative filter probability $\prod_{p}\sin^2(\theta_p)$ for QSITE, together with the single-measurement filter probabilities of QSVT/PrepSelPrep and QSVT/Embedding after real-part extraction; all three coincide at $0.469$ because at $t_{E_0}$ the embedding normalization $\alpha=\max_j|s_{E_j}|$ equals $\lambda=2$. \emph{Right:} single-shot probability of failure $P_F$. QSITE and the correctly extracted (real-part) QSVT state are identical ($P_F=0.533$); retaining the full complex QSVT output, including the imaginary leakage, degrades it to $P_F=0.665$. The exact ITE reference is shown for comparison ($P_F=0.034$).} 
    \label{fig:qsvt-vs-qsite}
\end{figure}

Figure~\ref{fig:qsvt-vs-qsite} illustrates the state-level equivalence for a representative 6-qubit instance: QSITE and the real-part-extracted QSVT output share the filter probability $0.469$ (the cumulative QSITE value) and the single-shot $P_F = 0.533$. Retaining the full complex QSVT output (that is, including the imaginary leakage discussed in Sec.~\ref{sec:qsvt-impl}) instead degrades the failure probability to $P_F = 0.665$, while the exact ITE reference gives $P_F = 0.034$.

\begin{figure}[H]
    \centering
    \includegraphics[width=0.56\linewidth]{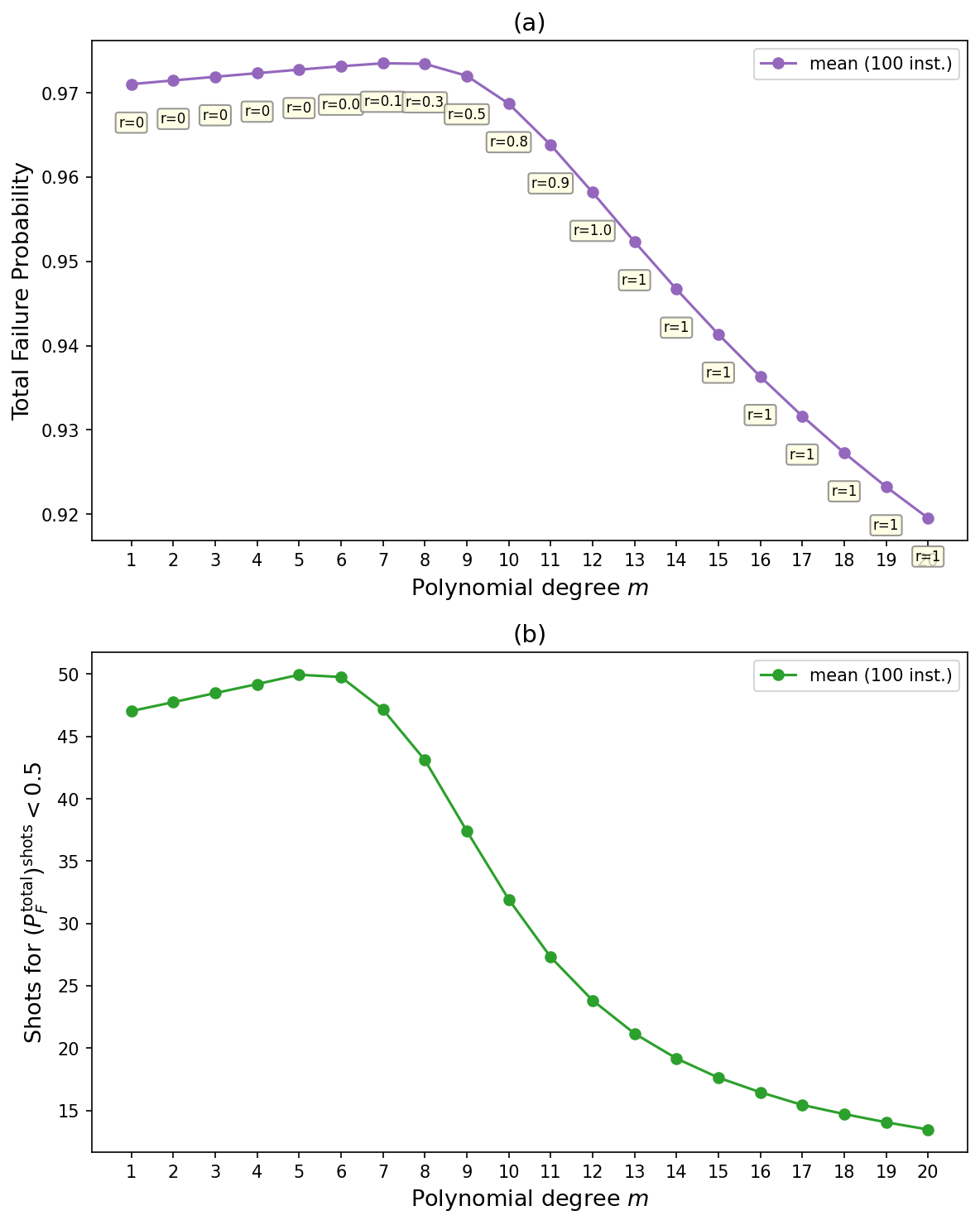}
	  \caption{Total failure probability $P_F^\mathrm{total}$ (a) and required number of shots (b) as a function of the polynomial degree $m$ for a 20-qubit MaxCut ensemble (100 instances, $k=20$, $\delta E = 8 \approx 12\%\,|E_0|$, $t=t_{E_0}$, QSVT + amplitude amplification). Amplification, with the smallest $r$ that raises the cumulative filter probability to at least $0.5$ (mean $r$ annotated), sets in from $m\approx9$; from there on, in contrast to the sequential schedule of Fig.~\ref{fig:joint-20q}, both quantities decrease with $m$.} 
    \label{fig:joint-20q-qsvt-aa}
\end{figure}

The practical payoff of the coherent circuit is amplitude amplification. In the first safe zone, the decreasing cumulative filter probability slightly outweighs the improving $P_F$ of the sequential schedule, so additional steps do not lower its total failure probability (Fig.~\ref{fig:joint-20q}). The coherent QSVT circuit lifts this limitation. Applying Grover amplitude amplification to this circuit, with the smallest number of iterations $r$ that raises the cumulative filter probability $\sin^2\theta=\prod_p\sin^2(\theta_p)$ to $\sin^2\big((2r+1)\theta\big)\geq 0.5$ (Sec.~\ref{sec:seq-qsite}), pins the sampling cost of the filter and leaves the total failure probability controlled by $P_F$, which keeps decreasing with $m$. Figure~\ref{fig:joint-20q-qsvt-aa} shows this for the 20-qubit ensemble at the first safe zone: amplification sets in once the cumulative filter probability drops below $0.5$ (from $m\approx9$), and from there on, in contrast to the sequential schedule of Fig.~\ref{fig:joint-20q}, both the total failure probability and the required shots decrease with $m$. 

\begin{figure}[H]
    \centering
    \includegraphics[width=0.55\linewidth]{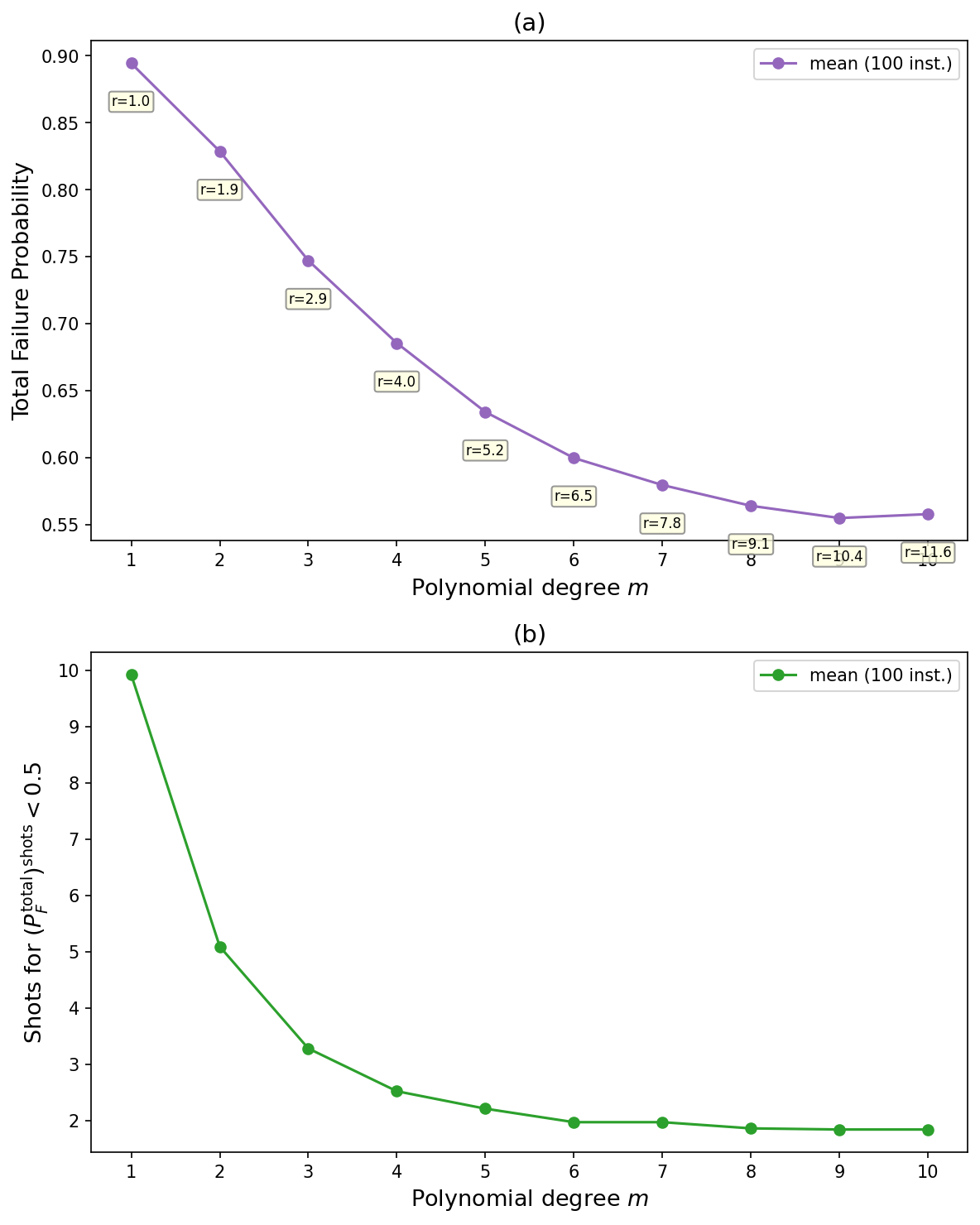}
    \caption{QSVT implementation of $S_k^m$ with Grover amplitude amplification for the 20-qubit MaxCut ensemble of Fig.~\ref{fig:grid-zone2-20q} (100 instances, $\delta E = 8$, $k=20$, second safe zone). (a)~Total failure probability $P_F^\mathrm{total}$ vs.\ polynomial degree $m$, with the smallest $r$ that raises the cumulative filter probability (Fig.~\ref{fig:grid-zone2-20q}(c)) to at least $0.5$ (mean $r$ annotated). (b)~Number of shots required for the total failure probability to drop below $0.5$.} 
    \label{fig:qsvt-aa-zone2-20q}
\end{figure}

At the second safe zone the cumulative filter probability is far smaller (Fig.~\ref{fig:grid-zone2-20q}(c)), and amplitude amplification is essential rather than merely advantageous. Figure~\ref{fig:qsvt-aa-zone2-20q} applies the same construction, with the same rule for $r$, to the same 20-qubit instances. The total failure probability decreases with $m$ and levels off near $0.55$, while the number of shots required to bring it below $0.5$ drops from about $10$ at $m=1$ to fewer than $2$ for $m \gtrsim 6$. The price of the amplification is circuit depth: the mean Grover iteration count grows roughly linearly with $m$ (from $\bar{r} \approx 1$ at $m=1$ to $\bar{r} \approx 12$ at $m=10$), each iteration invoking the degree-$m$ circuit and its inverse.

In both safe zones, amplitude amplification thus decouples the number of shots from the cumulative filter probability, which the sequential schedule cannot do.

\subsection{Shots and Number of Qubits}

\begin{figure}[H]
    \centering
    \includegraphics[width=0.70\linewidth]{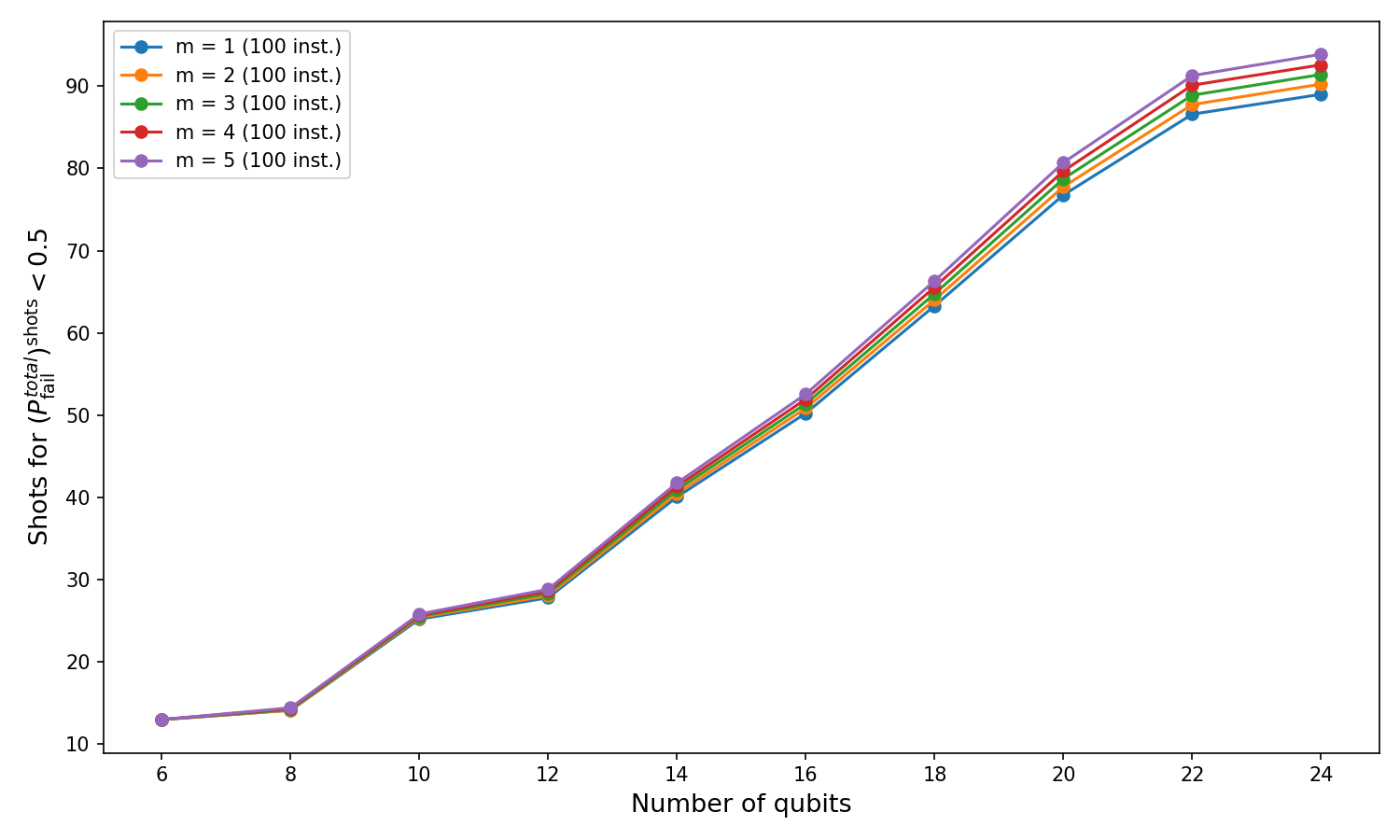}
    \caption{Number of shots required to achieve total failure probability $P_F^\mathrm{total} < 0.5$ as a function of the number of qubits $N$, for $m=1,2,3,4,5$ concatenation steps of the sequential QSITE schedule ($k=6$, first safe zone, $\delta E = 12\%\,|E_0|$, without amplitude amplification). Each point is the mean over 100 random MaxCut instances. The five curves nearly overlap (the improving $P_F$ and the decreasing cumulative filter probability nearly cancel), and $m=1$ needs the fewest shots. Results for $\delta E = 15\%\,|E_0|$ are in Fig.~\ref{fig:shots-vs-qubits-15pct}.}
    \label{fig:shots-vs-qubits}
    \label{fig:shots-vs-qubits-12pct}
\end{figure}

This subsection turns from fixed-size instances to how the sampling cost scales with the system size $N$. For the sequential schedule we report the number of shots $S$ needed to bring the total failure probability below $0.5$; for the coherent QSVT realization with Grover amplitude amplification (QSVT+AA) we plot the per-shot total failure probability $P_F^\mathrm{total}$, from which $S=\lceil \log 0.5/\log P_F^\mathrm{total}\rceil$ follows. Since $P_F^\mathrm{total}$ of the sequential schedule grows with $m$ (Fig.~\ref{fig:joint-20q}), we consider only small $m$. The main text shows the sequential schedule in the first safe zone (Fig.~\ref{fig:shots-vs-qubits}) and QSVT+AA in the second (Fig.~\ref{fig:qsvt-aa-zone2-vs-qubits}); the other two combinations are in Appendix~\ref{app:supplementary-shots}, and the dependence on $\delta E$ in Appendix~\ref{app:delta-e-sensitivity}.

We begin with the sequential QSITE schedule in the first safe zone. Figure~\ref{fig:shots-vs-qubits} shows the required shots as a function of $N$ (from $N=6$ to $N=24$) for $m=1,\dots,5$ at $t_{E_0}$, without amplitude amplification and with $\delta E = 12\%\,|E_0|$; each point is the mean over 100 instances. The five curves nearly overlap: as in Fig.~\ref{fig:joint-20q}, the improvement in $P_F$ is nearly cancelled, and slightly outweighed, by the decrease in cumulative filter probability. The shallowest circuit, $m=1$, therefore needs the fewest shots, and the shot count grows only moderately with system size (from ${\sim}\,13$ at $N=6$ to ${\sim}\,90$ at $N=24$). The required shot count is, however, sensitive to $\delta E$. Because MaxCut level degeneracies grow rapidly with the distance from the ground state, raising the tolerance from $12\%$ to $15\%$ of $|E_0|$ lowers the required shots by a factor that grows with $N$, from at most $2.5$ for $N\le14$ to about $4$ for $N\ge20$ (Appendix~\ref{app:delta-e-sensitivity}, Fig.~\ref{fig:shots-vs-qubits-15pct}).

It is instructive to disentangle the two competing effects by counting shots from the probability of failure alone, ignoring filter failures. Figure~\ref{fig:shots-pf-only} in Appendix~\ref{app:shots-pf-only} shows the number of shots $S$ needed so that $(P_F)^S<0.5$ as a function of $N$ for $m=5,10,20$ ($k=6$, first safe zone, $\delta E=12\%\,|E_0|$). Proposition~\ref{eq:pfsitemaxcutlinearinN} guarantees $P_F\le1/2$ with a single shot for $m=O(N)$ when $\delta E$ is a fixed fraction of $|E_0|$; with fewer steps the per-shot $P_F$ grows with $N$, but $(P_F)^S$ can still be driven below the target by increasing $S$. Counted this way, concatenation steps and shots are complementary resources: at $N=24$, $m=5$ requires about $2.3$ times more shots than $m=20$. Once filter failures are included, however, the shots grow slightly with $m$ (Fig.~\ref{fig:shots-vs-qubits}).

At the second safe zone the sequential schedule reaches a lower $P_F$ but a much smaller cumulative filter probability, so it needs more shots, and the shots grow steeply with $m$: at $N=24$, from ${\sim}\,120$ at $m=1$ to ${\sim}\,440$ at $m=5$ (Fig.~\ref{fig:shots-vs-qubits-secondzone}, Appendix~\ref{app:supplementary-shots}), against ${\sim}\,90$ in the first zone. For the sequential schedule the better state of the second zone is thus more than offset by the filter cost, which amplitude amplification of the coherent QSVT circuit can pin.

We now repeat the scaling analysis for QSVT+AA, with $k=N$, polynomial degrees $m=1,2,3$, 100 instances per $N$, and $\delta E$ chosen to be around $12\%\,|E_0|$. Per instance we apply $r=\lfloor \pi/(4\theta)-1/2\rfloor$ Grover iterations (the rule of Sec.~\ref{sec:seq-qsite}), where $\sin^2\theta$ is the cumulative filter probability, Eq.~\eqref{eq:cumulative-filter}, i.e.\ the success probability of the QSVT circuit after real-part extraction; the mean $r$ is annotated on the curves. In the first safe zone (Fig.~\ref{fig:qsvt-aa-zone1-vs-qubits}, Appendix~\ref{app:supplementary-shots}), $\sin^2\theta>1/4$ for every instance (mean $0.58$--$0.89$ over $N=6$--$14$), so $r=0$ and $P_F^\mathrm{total}$ is that of the sequential schedule: it rises from ${\sim}\,0.75$ at $N=6$ to ${\sim}\,0.95$ at $N=14$, and the three curves nearly coincide. At such shallow depth the bottleneck is $P_F$ itself, which so few concatenation steps barely reduce and which no amount of filter amplification can lower; deeper circuits can still improve $P_F^\mathrm{total}$ in the first zone (Fig.~\ref{fig:joint-20q-qsvt-aa}, $N=20$, $m\le20$). In the second safe zone (Fig.~\ref{fig:qsvt-aa-zone2-vs-qubits}) the success probability is small, so $r$ is nonzero and grows with $N$ and $m$ (for $m=3$, from ${\sim}\,1$ at $N=6$ to ${\sim}\,5$ at $N=20$). Because amplification pins the filter cost, deeper $m$ now helps: the curves separate and $m=3$ gives the lowest $P_F^\mathrm{total}$ at every size, ${\sim}\,0.63$ at $N=20$ versus ${\sim}\,0.90$ for $m=1$. The price is circuit depth, since each of the $r$ Grover iterations invokes the degree-$m$ circuit and its inverse.

  \begin{figure}[H]
    \centering
    \includegraphics[width=0.75\linewidth]{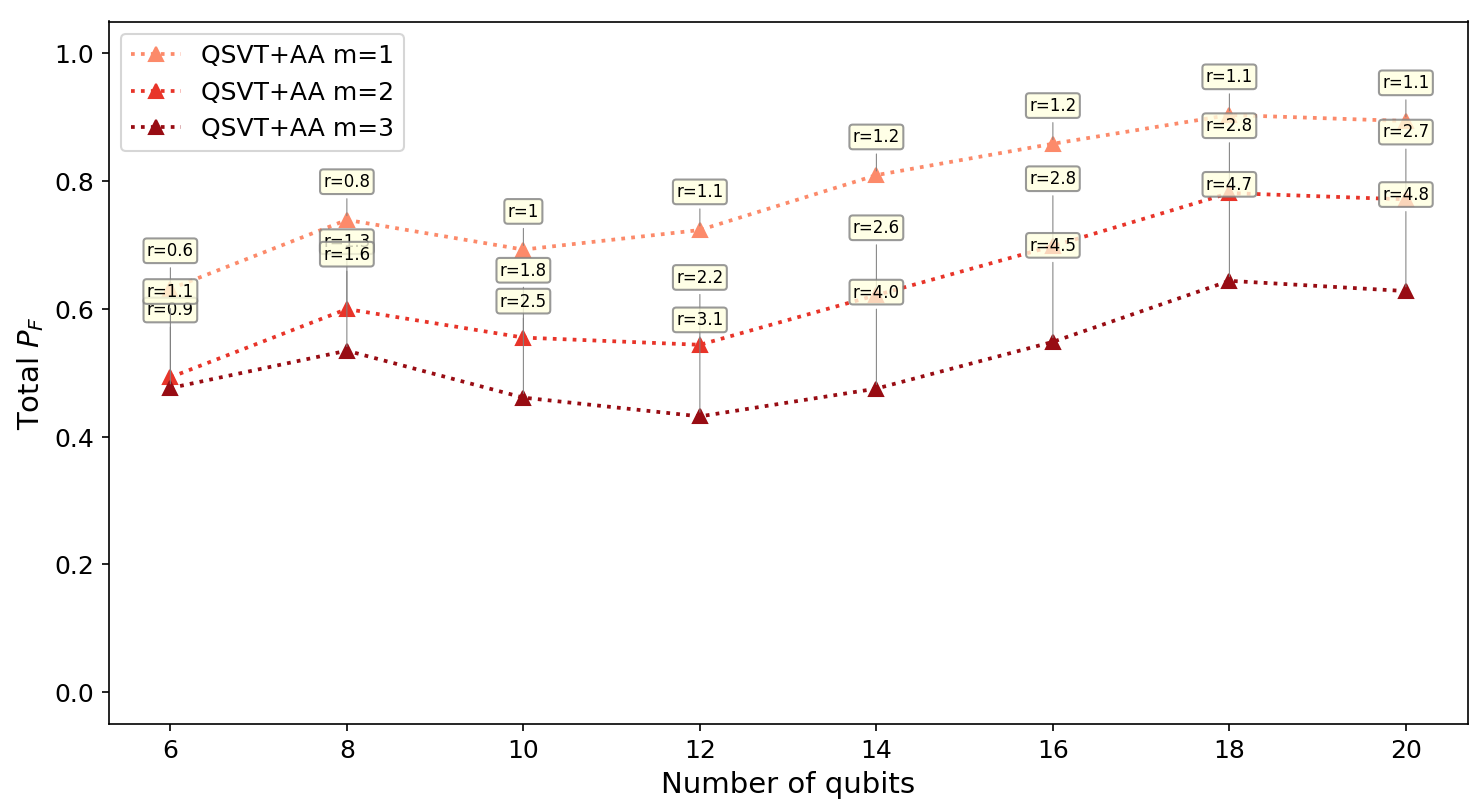}
	  \caption{Per-shot total failure probability $P_F^\mathrm{total}$ of the coherent QSVT realization with Grover amplitude amplification versus number of qubits $N$, at the \emph{second} safe zone ($k=N$, $m=1,2,3$, 100 random MaxCut instances, $\delta E$ around $12\%\,|E_0|$; shown up to $N=20$). The mean Grover count $r$ is annotated. Larger $m$ gives a lower $P_F^\mathrm{total}$ at every $N$ (${\sim}\,0.63$ at $N=20$ for $m=3$, versus ${\sim}\,0.90$ for $m=1$). This is the regime where QSVT+AA is advantageous over the sequential schedule; in the first safe zone $r=0$ (Fig.~\ref{fig:qsvt-aa-zone1-vs-qubits}, Appendix~\ref{app:supplementary-shots}).} 
    \label{fig:qsvt-aa-zone2-vs-qubits}
\end{figure}

In summary, for small $m$ the sequential schedule is best run in the first safe zone with $m=1$; there, amplitude amplification of the QSVT circuit is inactive ($r=0$) and offers no advantage. In the second safe zone the sequential shots grow steeply with $m$, whereas QSVT+AA turns larger $m$ into a lower $P_F^\mathrm{total}$. The corresponding number of shots $S$, averaged over the instances, is small: at $N=20$ about $9.5$, $4$ and $2$ for $m=1,2,3$, and below $3$ for $m=3$ over $N=6$--$20$. The second safe zone is thus useful only in combination with amplitude amplification. Beyond small $m$, at $N=20$, the total failure probability of sequential QSITE grows with $m$ (Fig.~\ref{fig:joint-20q}), whereas QSVT+AA lowers it in both zones (Figs.~\ref{fig:joint-20q-qsvt-aa} and~\ref{fig:qsvt-aa-zone2-20q}).

\subsection{Comparison with QAOA}\label{sec:qaoa-comparison}

A natural benchmark for optimization problems is the Quantum Approximate Optimization Algorithm (QAOA)~\cite{farhi2014quantumapproximateoptimizationalgorithm}. QAOA prepares a trial state by applying $p$ alternating layers of a cost unitary $e^{-i\gamma_l H_C}$ and a mixer unitary $e^{-i\beta_l H_M}$ (with $H_M = \sum_i X_i$) to the uniform superposition $|+\rangle^{\otimes N}$. The $2p$ variational angles $\{\gamma_l, \beta_l\}_{l=1}^{p}$ are then optimized classically to minimize $\langle H_C \rangle$. Unlike QSITE, whose parameters $t$, $k$, and $m$ are not optimized variationally, QAOA requires solving a non-convex classical optimization problem whose landscape grows in dimensionality with the circuit depth $p$. 

We compare QAOA and QSITE on the same ensemble of random MaxCut instances used throughout this work: Erd\"os--R\'enyi graphs $G(N, 1/2)$, 100 instances for each $N = 6, 8, 10, 12, 14$. The variational angles are optimized with COBYLA~\cite{Powell_1994_COBYLA}, from a near-zero initialization at $p=1$ and, at each $p>1$, as the better of two runs, one from a near-zero initialization and one from the INTERP warm-start heuristic of Zhou et al.~\cite{Zhou_2020}, which linearly interpolates the optimal angles from depth $p-1$.
Each evaluation of the QAOA cost function $\langle \psi(\boldsymbol{\gamma}, \boldsymbol{\beta}) | H_C | \psi(\boldsymbol{\gamma}, \boldsymbol{\beta}) \rangle$ requires executing the quantum circuit and performing measurements. These are shots used for training the QAOA circuit. Moreover, reaching depth $p$ requires the full optimization trajectory from $p=1$ through $p$ (since each depth is warm-started from the previous one), so the cost is \emph{accumulated} across all depths: $C_{\mathrm{QAOA}}(p) = \sum_{l=1}^{p} n_{\mathrm{iter}}(l)$, where $n_{\mathrm{iter}}(l)$ is the number of COBYLA evaluations in one run at depth $l$. Since two runs are made at each $l>1$, $C_{\mathrm{QAOA}}(p)$ is a lower bound on the number of circuit executions of QAOA.

Since QSITE does not have classical parameters to optimize, we start by comparing training shots of QAOA versus QSITE shots, allocating the budget $C_{\mathrm{QAOA}}(p)$ as the number of independent measurement shots for QSITE at concatenation depth $m = p$.

\begin{table}[H]
\centering
\begin{tabular}{c c c c c}
\toprule
$N$ & $n_{\mathrm{iter}}(p{=}1)$ & $n_{\mathrm{iter}}(p{=}2)$ & $n_{\mathrm{iter}}(p{=}3)$ & $C_{\mathrm{QAOA}}(p{=}3)$ \\
\midrule
6  & 10 & 20 & 30 & 60 \\
8  & 10 & 20 & 30 & 60 \\
10 & 20 & 30 & 40 & 90 \\
12 & 20 & 30 & 40 & 90 \\
14 & 20 & 40 & 70 & 130 \\
\bottomrule
\end{tabular}
\caption{COBYLA iteration budget per run at each QAOA depth and the accumulated budget $C_{\mathrm{QAOA}}$, which equals the number of QSITE shots allocated for comparison.} 
\label{tab:qaoa-maxiter}
\end{table}

\begin{figure}[H]
    \centering
    \includegraphics[width=0.75\linewidth]{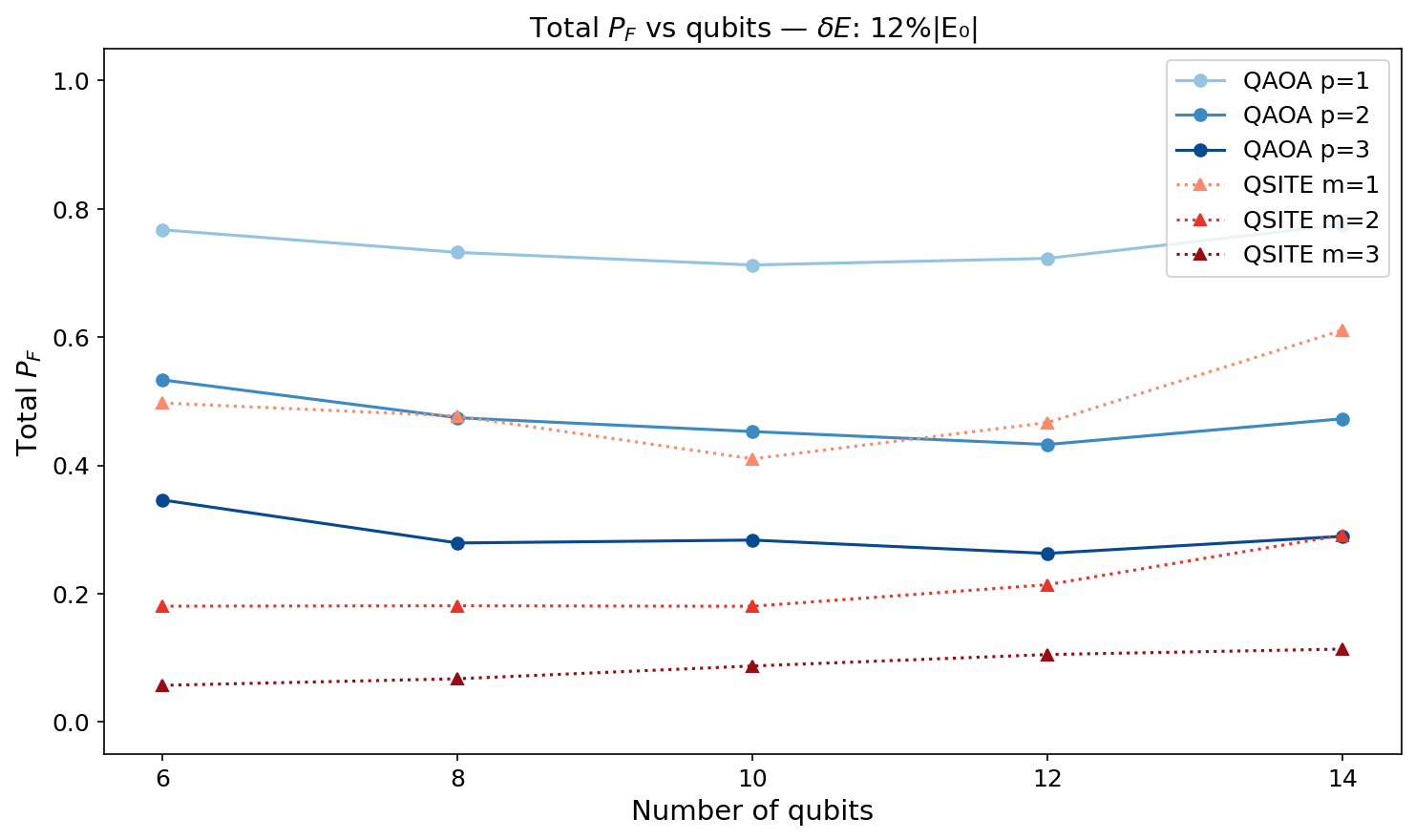}
    \caption{Total failure probability versus the number of qubits for QAOA (solid blue, depths $p=1,2,3$) and QSITE (dotted red, $m=1,2,3$ concatenations, $k=6$, first safe zone, no amplitude amplification), with $\delta E = 12\%\,|E_0|$ per instance; each point is the mean over 100 random MaxCut instances ($G(N, 1/2)$). QAOA: $P_F$ of one measurement of the trained state (Table~\ref{tab:qaoa-maxiter}). QSITE: probability $(P_F^\mathrm{total})^S$ that all $S = C_{\mathrm{QAOA}}(m)$ shots fail, which lies below QAOA at every depth and system size.}
    \label{fig:qaoa-vs-qsite}
\end{figure}

Figure~\ref{fig:qaoa-vs-qsite} shows the total failure probability as a function of the number of qubits for QAOA at depths $p = 1, 2, 3$ and QSITE at $m = 1, 2, 3$ concatenation steps in the first safe zone with no amplitude amplification. For QAOA, the total failure probability coincides with $P_F$ since there is no ancilla filter. For QSITE, the plotted quantity is $(P_F^\mathrm{total})^S$, averaged over 100 instances, where $S = C_{\mathrm{QAOA}}(m)$ shots are allocated according to Table~\ref{tab:qaoa-maxiter}. 

At $N = 14$, QAOA at $p = 3$ reaches $P_F \approx 0.29$, while QSITE at $m = 3$ with $130$ shots reaches ${\sim}\,0.11$, a factor of $2.5$ lower.
Each QSITE circuit execution is an independent measurement shot that returns an acceptable solution with probability $1 - P_F^\mathrm{total}$. Although this single-shot $P_F^\mathrm{total}$ degrades gradually with system size, $(P_F^\mathrm{total})^S$ can be driven to any target by increasing $S$ (Fig.~\ref{fig:shots-vs-qubits}).

However, this apparent advantage over QAOA only occurs at the training level. As soon as QAOA has finished training, its probability of failure is lower than the single-shot $P_F^\mathrm{total}$ of QSITE, so when measuring multiple times after training, its failure probability decreases much faster than that of QSITE with the same extra shots. On a real device, one would expect that each COBYLA iteration would itself require thousands of measurement shots to estimate $\langle H_C \rangle$ with sufficient precision, multiplying the true circuit budget by a large factor~\cite{Guerreschi_2019}.

In the second safe zone (QSVT+AA, Fig.~\ref{fig:qsvt-aa-zone2-vs-qubits}), the single-shot $P_F^\mathrm{total}$ of QSITE is comparable to the failure probability of QAOA after training at $m=p=1$, and at most $0.3$ higher at $m=p=2,3$. After training, QSITE thus needs at most a few times as many shots as QAOA to reach the same failure probability, whereas in the first safe zone, where a single QSITE shot succeeds far less often than trained QAOA, it needs far more.

\section{Conclusions}\label{sec:conclusions}

We have introduced QSITE, a quantum algorithm that approximates ITE through a sine-based expansion realized as a concatenation of Linear Combination of Unitaries circuits, requiring only two ancilla qubits. We proved that the QSITE operator converges uniformly to the exact ITE operator as the number of concatenation steps $m=k$ increases (Theorem~\ref{teo:site-operator}), and that the corresponding quantum state converges to the ideal ITE state (Theorem~\ref{teo:site-state}). For MaxCut Hamiltonians, we showed that $m = O(N)$ concatenation steps suffice to bound the probability of failure below $1/2$ when $\delta E$ is a fixed fraction of $|E_0|$ (Proposition~\ref{eq:pfsitemaxcutlinearinN}), and we established that the per-step filter probability increases monotonically with each concatenation step (Proposition~\ref{eq:monotoniafiltro}), so each successive ancilla measurement is more likely to succeed, although the probability of passing all of them decreases with $m$. 

Extensive numerical simulations on random Erd\"os--R\'enyi MaxCut instances with up to $24$ qubits revealed several notable features. The probability of failure exhibits a periodic structure with safe zones where $P_F$ can be systematically reduced. Within the first safe zone, the improving $P_F$ of the sequential schedule is nearly cancelled, and slightly outweighed, by the decreasing cumulative filter probability, so that $m=1$ needs the fewest shots. The shot requirements scale moderately with system size and are sensitive to the tolerable error $\delta E$, owing to the rapidly growing degeneracy of low-lying MaxCut energy levels. A second safe zone provides access to lower $P_F$ values at the cost of a much lower cumulative filter probability, which amplitude amplification of the coherent QSVT circuit can raise. For fully connected graphs, the high ground-state degeneracy keeps the required shots below $2$ even at $20$ qubits. 

We also compared the sequential implementation of QSITE with the QSVT one: the two prepare the same state, but the sequential one reaches it with a fixed two-ancilla circuit and mid-circuit measurements, avoiding the classical phase-angle synthesis and the real-part extraction that the QSVT realization needs. In exchange, only the QSVT realization admits amplitude amplification, which, once it sets in, turns additional concatenation steps into lower total failure probabilities and fewer shots in both safe zones, at the price of a Grover iteration count that grows with $m$. In the second safe zone, fewer than $2$ shots suffice at $20$ qubits. Across system sizes, the sequential schedule needs far more shots in the second safe zone than in the first, and they grow steeply with $m$ (about $120$ to $440$ at $N=24$ for $m=1$ to $5$, against about $90$ in the first); with amplitude amplification, the QSVT realization gains nothing in the first safe zone at $m\le3$, where $r=0$, but in the second needs fewer than $3$ shots for $m=3$ over $N=6$--$20$. 

Compared with QAOA on the same MaxCut instances, QSITE requires no classical optimization, so all circuit executions serve as independent measurement shots, whereas QAOA must invest a substantial fraction of its circuit budget in training the variational parameters. In the first safe zone, this gives QSITE a clear advantage during the training budget. In the second safe zone, the single-shot failure probability of QSITE with amplitude amplification is comparable to or moderately higher than that of trained QAOA, so the post-training advantage of QAOA is much smaller than in the first safe zone. 

Several directions remain open. The theoretical structure of the safe zones beyond $t_{E_0}$ deserves further analytical study, as does the regime $m > k$, which lies outside the domain of the main convergence theorems. Testing QSITE on Hamiltonians beyond MaxCut---such as maximum independent set or general QUBO problems---would clarify the generality of the shot-count scaling observed here. The performance of QSITE for generic, non-diagonal Hamiltonians and its precise relation to PITE~\cite{PhysRevResearch.4.033121},  also remain to be analyzed. Finally, noise model simulations and implementations on near-term quantum hardware would assess the algorithm's practical viability on current devices. 

\begin{acknowledgments}
This work was supported by Universidad Nacional de R\'io Negro [grant number \mbox{PI-UNRN} 2025 \mbox{40-B-1289}]
\end{acknowledgments}

\section*{Author Contributions}
All authors contributed equally.

During the preparation of this work the authors used Claude (Anthropic) to assist with drafting and editing parts of the text, preparing the figures, and checking computations in the examples. The authors reviewed and edited the content as needed and take full responsibility for the content of the published article.

\appendix

\section{Auxiliary lemmas and proofs of main results}
\label{app:aux_proofs}
In this appendix, we collect the proofs of the main results of this work, together with the statements and proofs of the auxiliary propositions and lemmas used throughout.

\subsection{Auxiliary lemmas and propositions}
	\begin{lemma} \label{propi sucesiones1}
		Let $c>0$, $t\in[0,c]$ and $H$ be a non-zero Hermitian matrix of $d\times d$, with eigenvalues
		\[E_0\leq E_1\leq...\leq E_{d-1},\]
		and let $m_0=\lceil c\|H\|\rceil\in\mathbb{N}$, which does not depend on $t\in[0,c]$. Then, for every $0\leq i\leq d-1$,
		\begin{enumerate}
			\item $\left\lbrace \left(1-\frac{E_it}{m} \right)^m  \right\rbrace_{m\in \mathbb{N}}$ converges uniformly to $e^{-E_it}$ in $[0,c]$.
			\item $\left\lbrace \left(1-\frac{E_it}{m} \right)^m  \right\rbrace_{m\geq m_0}$ is monotone non-decreasing.
			\item $0\leq\left(1-\frac{E_i t}{m}\right)^m\leq e^{\|H\|c}$ for every $m\geq m_0$.
			\item $0\leq\left(1-\sin\left(\frac{E_i t}{m}\right)\right)^m\leq e^{\|H\|c}$ for every $m\geq m_0$.
		\end{enumerate}
	\end{lemma}
	\begin{proof}
The first three items are standard properties of the sequence
\[
\left\{\left(1-\frac{x}{m}\right)^m\right\}_{m\in\mathbb{N}},
\]
with $x=E_it$ and $0\le t\le c$. It therefore remains to prove item~4: if $E_i=0$ or $t=0$, $1-\sin\left(\frac{E_i t}{m}\right)=1$ for every $m\in \mathbb{N}$. Suppose now that $E_i\neq 0$ and $t>0$. Let $m\geq m_0$ and observe that $\left|\frac{E_it}{m}\right|\leq 1<\frac{\pi}{2}$ for every $E_i$ and every $t\in [0,c]$: 
		\begin{itemize}
			\item if $E_i>0$, $1-\sin\left(\frac{E_i t}{m}\right)<1\Rightarrow \left(1-\sin\left(\frac{E_i t}{m}\right)\right)^m\leq 1=e^0<e^{\|H\|c}.$
			\item 	If $E_i<0$, $1-\sin\left(\frac{E_i t}{m}\right)=1+\sin\left(\frac{|E_i| t}{m}\right)$.
			By Taylor Theorem, $\sin\left(\frac{|E_i| t}{m}\right)=\frac{|E_i|}{m}t+R_1(t)$, where $R_1(t)=-\left( \frac{|E_i|}{m}\right)^2\sin\left(\frac{|E_i| \eta}{m}\right)\frac{t^2}{2}$ and $0<\eta<t$. Note that $R_1(t)<0$, since $t\in[0,c]$, $m\geq m_0$ and  $0\leq \frac{|E_i| \eta}{m}\leq \frac{|E_i| t}{m}<\frac{\pi}{2}$. Then,  
			\begin{align*}
				1+\sin\left(\frac{|E_i| t}{m}\right) & = 1+\frac{|E_i|}{m}t-\left( \frac{|E_i|}{m}\right)^2\sin\left(\frac{|E_i| \eta}{m}\right)\frac{t^2}{2}\leq 1+\frac{|E_i|}{m}t.
			\end{align*}
			Therefore,
			\[0\leq \left( 1-\sin\left(\frac{E_i t}{m}\right)\right) ^m=\left(1+\sin\left(\frac{|E_i| t}{m}\right)  \right)^m\leq \left(1+\frac{|E_i|}{m}t\right)^m\leq e^{|E_i|t}\leq e^{\|H\|c} .\]
		\end{itemize}
		So, we conclude that $0\leq\left( 1-\sin\left(\frac{E_i t}{m}\right)\right) ^m\leq e^{\|H\|c}$ for every $m\geq m_0$ and every $0\leq i\leq d-1$.

	\end{proof}

    	\begin{proposition} \label{propi sucesiones2}
		Let $c>0$, $t\in[0,c]$ and $H$ be a non-zero Hermitian matrix of $d\times d$, with eigenvalues 
		\[E_0\leq E_1\leq...\leq E_{d-1}.\]
		Then, the sequence $\left\{\left(1-\sin\left(\frac{E_i t}{m}\right)\right)^m\right\}_{m\in \mathbb{N}}$ converges uniformly to $e^{-E_it}$ in $[0,c]$ for every $0\leq i\leq d-1$.

	\end{proposition}

	\begin{proof}
Let $i\in\{0,1,...,d-1\}$ and $m\in \mathbb{N}$. Given real numbers $x,y$ and $m\in \mathbb{N}$, consider the well-known expansion
\begin{equation}\label{xn-yn}
	x^m-y^m= (x-y)\sum_{k=0}^{m-1}x^ky^{m-k-1}.
\end{equation}
If we replace $x=\left(1-\frac{E_i t}{m}\right)$ and $y=\left(1-\sin\left(\frac{E_i t}{m}\right)\right)$ in \eqref{xn-yn}, then
			\[\left(1-\frac{E_i t}{m}\right)^m-\left(1-\sin\left(\frac{E_i t}{m}\right)\right)^m= \left(1-\frac{E_i t}{m}-\left(1-\sin\left(\frac{E_i t}{m}\right)\right)\right)\sum_{k=0}^{m-1}\left(1-\frac{E_i t}{m}\right)^k\left(1-\sin\left(\frac{E_i t}{m}\right)\right)^{m-k-1}.\]
			Observe that $1-\sin\left(\frac{E_i t}{m}\right)\geq 0$ for every $m\in \mathbb{N}$. Taking modulus 
			\[\left|\left(1-\frac{E_i t}{m}\right)^m-\left(1-\sin\left(\frac{E_i t}{m}\right)\right)^m\right|\leq \left|-\frac{E_i t}{m}+\sin\left(\frac{E_i t}{m}\right)\right|\sum_{k=0}^{m-1}\left|1-\frac{E_i t}{m}\right|^{k}\cdot\left(1-\sin\left(\frac{E_i t}{m}\right)\right)^{m-k-1} \]
			By Taylor Theorem of order $2$, 
			\[\sin\left(\frac{E_i}{m}t\right)=\frac{E_i}{m}t+R_2,\]
			with $R_2(t)=-\left(\frac{E_i}{m}\right)^3\cos\left(\frac{\eta}{m}E_i\right)\frac{t^3}{3!}$ for some $\eta\in(0,t)$.
			Then, for every  $t\in[0,c]$.
			\begin{align*}
				\left|\left(1-\frac{E_i t}{m}\right)^m-\left(1-\sin\left(\frac{E_i t}{m}\right)\right)^m\right|&\leq \left|-\frac{E_i t}{m}+\frac{E_i}{m}t+R_2(t)\right|\sum_{k=0}^{m-1}\left|1-\frac{E_i t}{m}\right|^{k}\cdot\left(1-\sin\left(\frac{E_i t}{m}\right)\right)^{m-k-1} \\
				&\leq |R_2(t)|\sum_{k=0}^{m-1}K_{m,i}^{k}\cdot K_{m,i}^{m-k-1}= \left|\left(\frac{E_i}{m}\right)^3\cos\left(\frac{\eta}{m}E_i\right)\frac{t^3}{3!}\right|K_{m,i}^{m-1}\cdot m\\
				&\leq \frac {K_{m,i}^{m-1}}{6m^2}|E_it|^3\leq \frac {K_{m,i}^{m-1}}{6m^2}(\|H\|c)^3 ,
			\end{align*}
			where 
			\begin{equation}
				K_{m,i}= \max\left\{\left|1-\frac{E_i t}{m}\right|;1-\sin\left(\frac{E_i t}{m}\right)\right\}.
			\end{equation}
			Now consider $m\geq 2c\|H\|$ and note that $K_{m,i}^m\leq e^{\|H\|c}$ (by Lemma \ref{propi sucesiones1}, since $m\geq m_0$). Also $\frac{|E_i|t}{m}\leq \frac 12$, so
			\[K_{m,i}\geq\left|1-\frac{E_i t}{m}\right|\geq 1-\frac{|E_i|t}{m}\geq \frac 12. \]
			This means that
			\begin{align*}
				\left|\left(1-\frac{E_i t}{m}\right)^m-\left(1-\sin\left(\frac{E_i t}{m}\right)\right)^m\right|&\leq \frac {K_{m,i}^{m-1}}{6m^2}(\|H\|c)^3= \frac {K_{m,i}^{m}}{6m^2K_{m,i}}(\|H\|c)^3\\
				&\leq \frac{e^{\|H\|c}}{3m^2}(\|H\|c)^3
			\end{align*}
			for $m\geq 2c\|H\|$. Let $\varepsilon>0$,
			\begin{align*}
				\left|e^{-E_it}-\left(1-\sin\left(\frac{E_i t}{m}\right)\right)^m\right|&\leq  \left|e^{-E_it}-\left(1-\frac{E_i t}{m}\right)^m\right|+ \left|\left(1-\frac{E_i t}{m}\right)^m-\left(1-\sin\left(\frac{E_i t}{m}\right)\right)^m\right| \\
				&<\frac{\varepsilon}{2}+\frac{e^{\|H\|c}}{3m^2}(\|H\|c)^3<\frac{\varepsilon}{2}+\frac{\varepsilon}{2}=\varepsilon
			\end{align*}
			for every
			\begin{equation}\label{m0 para convergencia unif de site}
				m> \max\left\lbrace M_{1,i}; 2c\|H\|;\sqrt{\frac{2e^{\|H\|c}(\|H\|c)^3}{3\varepsilon}}\right\rbrace,
			\end{equation}
            where $M_{1,i}\in\mathbb{N}$ is such that $\left|e^{-E_it}-\left(1-\frac{E_i t}{m}\right)^m\right|<\frac{\varepsilon}{2}$ for every $m\geq M_{1,i}$ and every $t\in[0,c]$ (it exists by item 1 of Lemma \ref{propi sucesiones1}).
			Observe that if $m$ satisfies \eqref{m0 para convergencia unif de site}, then $m\geq m_0$ and fulfills all items in Lemma \ref{propi sucesiones1}. 
			Since \eqref{m0 para convergencia unif de site} does not depend on the value of $t$ in the interval $[0,c]$, we can conclude that $\left\lbrace \left(1-\sin\left(\frac{E_i t}{m}\right)\right)^m\right\rbrace_{m\in \mathbb{N}}$ converges uniformly to $e^{-E_it}$ in $[0,c]$ 
			for every $0\leq i\leq d-1$.

	\end{proof}

	\begin{proposition}\label{coro2}
		Let $t\in [0,c]$ and $H$ be a Hermitian matrix of $d\times d$, with eigenvalues
		\[E_0\leq E_1\leq...\leq E_{d-1}.\]
		Then, $f(tH)=Uf(tD)U^{\dagger}$	for every continuous function $f:\mathbb{R}\to\mathbb{R}$ and, for any initial state $|\psi^0\rangle$
		\begin{enumerate}
			\item $e^{-tH}\ket{\psi^0}=\sum_{i=0}^{d-1}e^{-tE_i}\langle\psi^0|E_i\rangle |E_i\rangle$.
			\item $\left(I-\frac{tH}{m}\right)^m\ket{\psi^0}=\sum_{i=0}^{d-1}\left(1-\frac{tE_i}{m}\right)^m\langle\psi^0|E_i\rangle |E_i\rangle$.
			\item $\left(I+ \dfrac{e^{i\frac{-tH}{m}}-e^{i\frac{tH}{m}}}{2i}\right)^m|\psi^0\rangle= \sum_{i=0}^{d-1}s_{E_i}^m\langle\psi^0|E_i\rangle |E_i\rangle$ with $s_{E_i}^m=\left(1-\sin\left(\frac{E_it}{m}\right)\right)^m$, for every $0\leq i\leq d-1$.
			\item $\left(I+ \dfrac{e^{i\frac{-tH}{m}}-e^{i\frac{tH}{m}}}{2i}\right)^m|\psi^0\rangle= 0$ if and only if $|\psi^0\rangle\perp|E_i\rangle$ for every $i$ such that $s_{E_i}\neq 0$.
			\item The operator $\left(I+ \dfrac{e^{i\frac{-tH}{m}}-e^{i\frac{tH}{m}}}{2i}\right)^m$ has trivial kernel for every $m\geq 2c\|H\|$. Moreover, in this case
			\[\left\|\left(I+ \dfrac{e^{i\frac{-tH}{m}}-e^{i\frac{tH}{m}}}{2i}\right)^m|\psi^0\rangle\right\|\neq 0.\]

		\end{enumerate}

	\end{proposition}
	
	\begin{proof}
		Since $H$ is Hermitian, it can be written as $UDU^{\dagger}$, with 
		\[U=\begin{pmatrix}
			|E_0\rangle|&|E_1\rangle|&...|&|E_{d-1}\rangle\\  
		\end{pmatrix}\quad \text{ and}\quad D=\begin{pmatrix}
			E_0&0&\cdots&0\\
			0&E_1&0\cdots&0\\
			\cdots&\cdots&\cdots&\cdots\\
			0&\cdots&0&E_{d-1}
		\end{pmatrix}.\]
		Then, by the spectral theorem $f(tH)=Uf(tD)U^{\dagger}$ and
		\begin{align*}
			f\left(Ht\right)|\psi^0\rangle&=\begin{pmatrix}
				|E_0\rangle|&|E_1\rangle|&...|&|E_{d-1}\rangle\\  
			\end{pmatrix}\begin{pmatrix}
				f\left(E_0t\right)&0&\cdots&0\\
				0&f\left(E_1t\right)&0\cdots&0\\
				\cdots&\cdots&\cdots&\cdots\\
				0&\cdots&0&f\left(E_{d-1}t\right)
			\end{pmatrix}\begin{pmatrix}
				\langle E_0|\\
				\langle E_1|\\
				\vdots\\
				\langle E_{d-1}|\\
			\end{pmatrix}\ket{\psi^0}\\
			&= f\left(E_0t\right)\langle\psi^0|E_0\rangle |E_0\rangle+f\left(E_1t\right)\langle\psi^0|E_1\rangle |E_1\rangle+\cdots+f\left(E_{d-1}t\right)\langle\psi^0|E_{d-1}\rangle |E_{d-1}\rangle.
		\end{align*}
Therefore, the explicit formulas in items 1--3 are obtained by taking
\[
f(x)=e^{-x},\qquad
f(x)=\left(1-\frac{x}{m}\right)^m,\qquad
f(x)=\left(1-\sin\left(\frac{x}{m}\right)\right)^m,
\]
respectively.

		On the other hand, to prove item 4, observe that $\left(I+ \dfrac{e^{i\frac{-tH}{m}}-e^{i\frac{tH}{m}}}{2i}\right)^m|\psi^0\rangle= 0$ if and only if 
		\[\sum_{i=0}^{d-1}s_{E_i}^m\langle\psi^0|E_i\rangle |E_i\rangle=0\Leftrightarrow s_{E_i}=0\ \lor \  \langle\psi^0|E_i\rangle=0 \]
		for each $0\leq i\leq d-1$. If $s_{E_i}\neq0,$ then $\ket{\psi^0}$ must be orthogonal to $\ket{E_i}$. In general, note that if there exists $E_i\neq mE_j+2km\pi$, then $s_{E_i}$ and $s_{E_j}$ cannot be simultaneously $0$ for the same $t$.
		
		To prove item 5, note that if $m\geq 2c\|H\|$, then 
		\[\left|\frac{E_it}{m}\right|\leq \frac{\|H\|t}{m}\leq \frac{\|H\|c}{m}\leq \frac 12< \frac{\pi}{4}  \]
		which means that $1-\sin\left(\frac{E_it}{m}\right)>0$
		for every $0\leq i\leq d-1$ and $t\in [0,c]$. Therefore, 
		\[\sum_{i=0}^{d-1}s_{E_i}^m\langle\psi^0|E_i\rangle |E_i\rangle= 0\Rightarrow \ket{\psi^0}\perp\ket{E_i}\ \forall\ i,\]
		so we conclude that there is no initial state $\ket{\psi^0}\neq 0$ such that $\left(I+ \dfrac{e^{i\frac{-tH}{m}}-e^{i\frac{tH}{m}}}{2i}\right)^m|\psi^0\rangle=\sum_{i=0}^{d-1}s_{E_i}^m\langle\psi^0|E_i\rangle |E_i\rangle= 0$ for $m\geq 2c\|H\|$.
		
	\end{proof}

\subsection{Proof of Thm. \eqref{teo:site-operator}}	

\begin{proof}	
In this proof we relabel $k$ by the concatenation label $m$ for convenience of reading. It suffices to prove the convergence in the spectral norm. Using the diagonalization of $H$ and Proposition \ref{coro2} 

\begin{align*}
	\left\|e^{-tH}-\left(I+ \dfrac{e^{i\frac{-tH}{m}}-e^{i\frac{tH}{m}}}{2i}\right)^m\right\|&=   \left\|Ue^{-tD}U^{\dagger}-\left(UU^{\dagger}+ \dfrac{Ue^{i\frac{-tD}{m}}U^{\dagger}-Ue^{i\frac{tD}{m}}U^{\dagger}}{2i}\right)^m\right\|\\
	&= \left\|Ue^{-tD}U^{\dagger}-U\left(I+ \dfrac{e^{i\frac{-tD}{m}}-e^{i\frac{tD}{m}}}{2i}\right)^mU^{\dagger}\right\|\\
	&= \left\|e^{-tD}-\left(I+ \dfrac{e^{i\frac{-tD}{m}}-e^{i\frac{tD}{m}}}{2i}\right)^m\right\|.
\end{align*}
Observe that $e^{-tD}-\left(I+ \dfrac{e^{i\frac{-tD}{m}}-e^{i\frac{tD}{m}}}{2i}\right)^m$ is a diagonal matrix whose entries are
\[e^{-tE_j}-\left(1+ \dfrac{e^{i\frac{-tE_j}{m}}-e^{i\frac{tE_j}{m}}}{2i}\right)^m,\ \]
for every $j=0,1,...,d-1$. Moreover, 
\[\dfrac{e^{\frac{-it}{m}E_j}-e^{\frac{it}{m}E_j}}{2i}=\sin\left(-\frac{t}{m}E_j\right).\]
Therefore, 
\begin{equation} \label{norma e-1+sen}
	\left\|e^{-tH}-\left(I+ \dfrac{e^{\frac{-it}{m}H}-e^{\frac{it}{m}H}}{2i}\right)^m\right\|= \max_{0\leq j\leq d-1}\left|e^{-tE_j}-\left(1-\sin\left(\frac{t}{m}E_j\right)\right)^m\right|
\end{equation}
Let $\varepsilon>0$. Lemma \ref{propi sucesiones1} and Proposition \ref{propi sucesiones2} allow us to set the following conditions:
	\begin{enumerate}
		\item There exists $M_1\in \mathbb{N}$ such that $\left|e^{-E_jt}-\left(1-\frac{E_jt}{m} \right)^m\right|<\frac{\varepsilon}{2}$ for every $m\geq M_1$, every $t\in[0,c]$ and every $0\leq j\leq d-1$.
		\item $M_0\in \mathbb{N}$ such that $M_0> \max\left\lbrace M_{1}; 2c\|H\|;\sqrt{\frac{2e^{\|H\|c}(\|H\|c)^3}{3\varepsilon}}\right\rbrace$.
	\end{enumerate}
	Then,
	\begin{equation}\label{conv M0}
		m\geq M_0\Rightarrow\ \left|e^{-tE_j}-\left(1-\sin\left(\frac{E_j t}{m}\right)\right)^m\right|<\varepsilon,
	\end{equation}
 for every $0\leq j\leq d-1$.
 Therefore, 
\[        \left\|e^{-tH}-\left(I+ \dfrac{e^{\frac{-it}{m}H}-e^{\frac{it}{m}H}}{2i}\right)^m\right\|= \max_{0\leq j\leq d-1}\left|e^{-tE_j}-\left(1-\sin\left(\frac{t}{m}E_j\right)\right)^m\right|<\varepsilon \]
and $\left(I+ \dfrac{e^{i\frac{-tH}{m}}-e^{i\frac{tH}{m}}}{2i}\right)^m\longrightarrow\ e^{-tH}$ in the spectral norm.

Observe that $M_0$ is independent of t on an interval of the form $[0,c]$, for some $c>0$. Hence, the operator sequence converges uniformly on this interval.

\end{proof}
	
\subsection{Proof of Thm. \eqref{teo:site-state}}	
	
	\begin{proof}
		In this proof we relabel $k$ by the concatenation label $m$ for convenience of reading. Given any initial state $\ket{\psi^0}$
		\begin{equation}
			\ket{\psi^{ite}_t} =  \dfrac{e^{-tH}\ket{\psi^0}}{\|e^{-tH}\ket{\psi^0}\|}\quad \text{ and } 	\ket{\psi^{qsite}_{t, m}}= \dfrac{\left(I+ \dfrac{e^{i\frac{-tH}{m}}-e^{i\frac{tH}{m}}}{2i}\right)^m|\psi^0\rangle}{\left\|\left(I+ \dfrac{e^{i\frac{-tH}{m}}-e^{i\frac{tH}{m}}}{2i}\right)^m|\psi^0\rangle\right\|} 
		\end{equation}
		
	Observe that if $x$ and $y$ are unit vectors in a Hilbert space, then
	\[
	\|x-y\|^2=\|x\|^2+\|y\|^2-2\operatorname{Re}\langle x|y\rangle
	=2-2\operatorname{Re}\langle x|y\rangle\ge 0.
	\]
	Hence, $\operatorname{Re}\langle x \mid y\rangle \leq 1$. Applying this identity with
	$x=\ket{\psi^{ite}_{t}}$ and $y=\ket{\psi^{qsite}_{t,m}}$, we obtain 
	\begin{equation} \label{norm and inner}
		\left\|\ket{\psi^{ite}_{t}}-\ket{\psi^{qsite}_{t,m}}\right\|^2 
		=
		2-2\operatorname{Re}\left(\langle\psi^{ite}_{t}\mid\psi^{qsite}_{t,m}\rangle\right). 
	\end{equation}
	Consequently, our analysis reduces to estimating the inner product
	$\langle\psi^{ite}_{t}\mid\psi^{qsite}_{t,m}\rangle$. It is important to note that $\langle\psi^{ite}_{t}\mid\psi^{qsite}_{t,m}\rangle$ is real and non-negative. Moreover, 
		\[0\leq\langle\psi^{ite}_{t}\mid\psi^{qsite}_{t,m}\rangle\leq 1.\]

By the operator convergence in Theorem \ref{teo:site-operator} 
	\[\lim_{m\to \infty}\left\|e^{-tH}- \left(I+ \dfrac{e^{i\frac{-tH}{m}}-e^{i\frac{tH}{m}}}{2i}\right)^m\right\|=0,\]
	for $t\in [0,c]$. Then,
	\begin{align*}
		\lim_{m\to \infty}\left\|e^{-tH}|\psi^0\rangle- \left(I+ \dfrac{e^{i\frac{-tH}{m}}-e^{i\frac{tH}{m}}}{2i}\right)^m|\psi^0\rangle \right\|& \leq \lim_{m\to \infty} \left\|e^{-tH}- \left(I+ \dfrac{e^{i\frac{-tH}{m}}-e^{i\frac{tH}{m}}}{2i}\right)^m \right\|=0, 
	\end{align*}
	hence $\lim\limits_{m\to \infty}\left(I+ \dfrac{e^{i\frac{-tH}{m}}-e^{i\frac{tH}{m}}}{2i}\right)^m|\psi^0\rangle=e^{-tH}|\psi^0\rangle$. Also recall that $\left\|\left(I+ \dfrac{e^{i\frac{-tH}{m}}-e^{i\frac{tH}{m}}}{2i}\right)^m|\psi^0\rangle\right\|\neq 0$ for $m\geq M_0\geq 2c\|H\|$ (see Proposition \ref{coro2}). Therefore, using the continuity of the norm and the inner product follows that
	\begin{align*}
	\lim_{m\to \infty}  \left\langle \psi^{ite}_{t}\mid\psi^{qsite}_{t,m}\right\rangle 
&= \lim_{m\to \infty} \dfrac{\left\langle e^{-tH}\ket{\psi^0}\mid \left(I+ \dfrac{e^{i\frac{-tH}{m}}-e^{i\frac{tH}{m}}}{2i}\right)^m\ket{\psi^0}\right\rangle 	}{\|e^{-tH}\ket{\psi^0}\|\left\|\left(I+ \dfrac{e^{i\frac{-tH}{m}}-e^{i\frac{tH}{m}}}{2i}\right)^m\ket{\psi^0}\right\|}= \dfrac{\left\langle e^{-tH}\ket{\psi^0}\mid  e^{-tH}\ket{\psi^0}\right\rangle 	}{\|e^{-tH}\ket{\psi^0}\|\|e^{-tH}\ket{\psi^0}\|}=1.
	\end{align*}
	Finally, taking limit in \eqref{norm and inner}
	\[	\lim_{m\to \infty} 	\left\|\ket{\psi^{ite}_{t}}-\ket{\psi^{qsite}_{t,m}}\right\|^2 
	=\lim_{m\to \infty}
	2-2\operatorname{Re}\left(\langle\psi^{ite}_{t}\mid\psi^{qsite}_{t,m}\rangle\right)=0. 
	\]

	\end{proof}

\subsection{Proof of proposition \eqref{eq:monotoniafiltro}}
To see this let $|\psi^0\rangle\in \mathcal{H}$ be an initial state $\|\psi^0\|=1$ and we denote $S=I -\frac{ie^{-\frac{it}{m}H}}{2}+\frac{ie^{\frac{it}{m}H}}{2}$. Then $\sin(\theta_0)=\frac 12 \left\|S|\psi^0\rangle\right\|$.
We define as $R$ the orthogonal projection onto the subspace spanned by $|00\rangle\otimes|\psi_0\rangle$, that is
\[R|\Psi\rangle= \frac{|00\rangle}{2}\left[I -\frac{ie^{-\frac{it}{m}H}}{2}+\frac{ie^{\frac{it}{m}H}}{2} \right]|\psi^0\rangle= \frac{|00\rangle}{2}S|\psi^0\rangle\ \Rightarrow\dfrac{R|\Psi\rangle}{\|R|\Psi\rangle\|}=\frac{|00\rangle S|\psi^0\rangle}{\|S|\psi^0\rangle\|}.\]
Applying $I\otimes S$
\begin{equation}
	(I\otimes S)\left( \frac{|00\rangle S|\psi^0\rangle}{\|S|\psi^0\rangle\|}\right)= \frac{|00\rangle S^2|\psi^0\rangle}{\|S|\psi^0\rangle\|}=\frac{|00\rangle S^2|\psi^0\rangle}{2\sin(\theta_0)}.
\end{equation}
\begin{itemize}
	\item  If $|\psi^1\rangle=\dfrac{|00\rangle S|\psi^0\rangle}{2\sin(\theta_0)}$, then $\|\psi^1\|=1$ and
	\begin{equation}
		\sin(\theta_1)=\frac 12 \left\|(I\otimes S)|\psi^1\rangle\right\|= \frac{1}{4}\left\| S^2\frac{|\psi^0\rangle}{\sin(\theta_0)}\right\|.
	\end{equation}
	\item  If $|\psi^2\rangle=\dfrac{(I\otimes S)|\psi^1\rangle}{\|(I\otimes S)|\psi^1\rangle\|}=\dfrac{|00\rangle S^2|\psi^0\rangle}{4\sin(\theta_0)\sin(\theta_1)}$, then $\|\psi^2\|=1$ and
	\begin{equation}
	\sin(\theta_2)=\frac 12 \left\| (I\otimes S)|\psi^2\rangle\right\|= \frac{1}{8}\left\| S^3\frac{|\psi^0\rangle}{\sin(\theta_0)\sin(\theta_1)}\right\|.
	\end{equation}
	\item In general, for every $1\leq p\leq m-1$,
	\[|\psi^{p}\rangle=\dfrac{(I\otimes S)|\psi^{p-1}\rangle}{\|(I\otimes S)|\psi^{p-1}\rangle\|}=\dfrac{|00\rangle S^{p}|\psi^0\rangle}{2^p\prod_{l=0}^{p-1}\sin(\theta_l)},\]
	with $\|\psi^{p}\|=1$ and
	\begin{equation} \label{sen thetak}
		\sin(\theta_p)=\frac 12 \left\| (I\otimes S)|\psi^p\rangle\right\|= \frac{1}{2^{p+1}}\left\| S^{p+1}\frac{|\psi^0\rangle}{\prod_{l=0}^{p-1}\sin(\theta_l)}\right\|.
	\end{equation}
	Note that for every $p$, $\sin(\theta_p)\in [0,1]$ and
	\begin{equation}\label{norma S^k}
		\|S^{p+1}|\psi^0\rangle\|=2^{p+1}\prod_{l=0}^{p}\sin(\theta_l).
	\end{equation}

\end{itemize}

It remains to establish the monotonicity itself. Denote $a_p:=\|S^{p}|\psi^0\rangle\|$. Dividing two consecutive instances of Eq.~\eqref{norma S^k} telescopes the product and yields the ratio identity
\begin{equation}\label{eq:ratio-identity}
    \sin(\theta_p)=\frac 12\,\frac{a_{p+1}}{a_p}=\frac 12\,\frac{\|S^{p+1}|\psi^0\rangle\|}{\|S^{p}|\psi^0\rangle\|},
\end{equation}
valid whenever $\sin(\theta_l)>0$ for $l<p$, i.e.\ precisely in the regime where the concatenation is defined. With Eq.~\eqref{eq:ratio-identity}, the claim $\sin(\theta_{p-1})\leq\sin(\theta_p)$ is equivalent to
\begin{equation}\label{eq:log-convexity}
    \frac{a_p}{a_{p-1}}\leq\frac{a_{p+1}}{a_p}\qquad\Longleftrightarrow\qquad a_p^2\leq a_{p-1}\,a_{p+1},
\end{equation}
that is, to the log-convexity of the sequence $(a_p)_p$.

The single ingredient needed is that $S$ is Hermitian: since $H=H^\dagger$,
\[
    S^\dagger=I+\frac{i}{2}e^{\frac{it}{m}H}-\frac{i}{2}e^{-\frac{it}{m}H}=S,
\]
equivalently $S=I-\sin\!\left(\tfrac{t}{m}H\right)$, a real function of $H$. Using $S^\dagger=S$ we may slide one factor of $S$ across the inner product,
\[
    a_p^2=\langle S^{p}\psi^0\,|\,S^{p}\psi^0\rangle=\langle \psi^0\,|\,S^{2p}\,|\,\psi^0\rangle=\langle S^{p-1}\psi^0\,|\,S^{p+1}\psi^0\rangle,
\]
and Cauchy--Schwarz applied to the last inner product gives
\[
    a_p^2=\langle S^{p-1}\psi^0\,|\,S^{p+1}\psi^0\rangle\leq \|S^{p-1}\psi^0\|\,\|S^{p+1}\psi^0\|=a_{p-1}\,a_{p+1},
\]
which is exactly Eq.~\eqref{eq:log-convexity}. Hence the ratios $a_{p+1}/a_p$ are non-decreasing and, by Eq.~\eqref{eq:ratio-identity}, $\sin(\theta_{p-1})\leq\sin(\theta_p)$ for every $1\leq p\leq m$. $\hfill\blacksquare$

\subsection{Proof of Lemma \eqref{lemma bound}}
\begin{proof}
\begin{enumerate}
    \item Expression \eqref{seno vectorial version} is proved in Proposition \ref{coro2}.
    \item Observe that $\sin(\theta_0)=0$ if and only if 
\begin{equation}\label{cero site}
   1-\sin\left(\frac{E_it}{k} \right)=0\quad \text{ or }\quad \langle\psi^0|E_i\rangle=0
\end{equation}
for each $0\leq i\leq d-1$. Suppose that $\sin(\theta_0)=0$, since $\ket{\psi^0}\neq 0$ there exists at least one $i_0$ such that $1-\sin\left(\frac{E_{i_0}t}{k} \right)=0$ and $\langle\psi^0|E_{i_0}\rangle\neq 0$, which means that $\frac{E_{i_0}t}{k}=\frac{\pi}{2}+2k'\pi$, $k'\in \mathbb{Z}$. If we readjust $k$ to $k+1$, then $\frac{E_{i_0}t}{k+1}<\frac{E_{i_0}t}{k}$, $1-\sin\left(\frac{E_{i_0}t}{k+1} \right)\neq0$ for the same $t$, and eventually we set 
\[    \sin(\theta_0) = \frac{\|S_{k+1}|\psi^0\rangle \|}{2}>0.
\]
For more details, see Proposition \ref{coro2}. 
    \item The expression \eqref{seno vectorial version} is related to the quadratic mean
\[\sin(\theta) =\frac{\sqrt{d}}{2}QM(x)\]
of $x= (x_1,x_2,...,x_{d})$, with 
\[ x_i= \left|1-\sin\left(\frac{E_it}{k}\right)\right||\langle\psi_0|E_i\rangle|, 1\leq i\leq d.\]
Since the Quadratic mean is greater than the Arithmetic mean $AM(x)=\frac 1d \sum_{i=0}^{d-1}x_i$, follows that
\begin{equation}\label{..}
    \sin(\theta_0)\geq \frac{\sqrt{d}}{2d}\sum_{i=0}^{d-1} \left(1-\sin\left(\frac{E_it}{k}\right)\right)|\langle\psi^0|E_i\rangle|.
\end{equation} 
Suppose now we take a special initial state such as $|\psi^0\rangle=\frac {1}{\sqrt{d}}\sum_{j=0}^{d-1}|E_i\rangle$ (which will be useful in optimization), then for $k>\frac{6\|H\|t}{\pi}$:

\begin{align}
    \sin(\theta_0)&\geq \frac {1}{2d}\sum_{i=0}^{d-1}\left(1-\sin\left(\frac{E_it}{k}\right)\right) \\
    & =\dfrac{d-\sum_{i=0}^{d-1}\sin\left(\frac{E_it}{k}\right)}{2d}\\    \nonumber
    &\geq \dfrac{d-\frac d2}{2d}=\frac{1}{4}. 
\end{align}

\end{enumerate}

\end{proof}

\subsection{Proof of Proposition \eqref{eq:pfsitemaxcutlinearinN}}	
	
	\begin{proof}
		Since $E_0\leq...\leq E_{d-1}$ and are all non-positive let us write $t_{E_0}=\frac{\pi}{2}\frac{k}{|E_0|}$: 
		\begin{equation}\label{eq:boundedPf_proof}
			P^{qsite}_{F,~k,m}(t_{E_0})=\frac{\sum^{d-1}_{i=r}s^{2m}_{-|E_{i}|,k}(t_{E_0})}{\sum^{r-1}_{i=0}s^{2m}_{-|E_{i}|,k}(t_{E_0})+\sum^{d-1}_{i=r}s^{2m}_{-|E_{i}|,k}(t_{E_0})}. 
		\end{equation}
		
		On one hand, the first term in the denominator is lower-bounded by $g 2^{2m} \leq \sum^{r-1}_{i=0}s^{2m}_{-|E_{i}|,k}(t_{E_0})$, where $g$ is the degeneracy of the ground state.
		On the other hand, $s^{2m}_{-|E_i|,k}(t_{E_0})\leq s^{2m}_{-|E_r|,k}(t_{E_0})$ for every $i\geq r$, since $|E_i|\leq|E_r|$, so $\sum^{d-1}_{i=r}s^{2m}_{-|E_{i}|,k}(t_{E_0})\leq (d-r)s^{2m}_{-|E_{r}|,k}(t_{E_0})$. Since the numerator and the second term of denominator of \eqref{eq:boundedPf_proof} are equal, $P^{qsite}_{F,~k,m}$ is a monotonic increasing function of this term. So we can use these bounds: 
		\begin{equation}
			P^{qsite}_{F,~k,m}(t_{E_0})\leq \frac{(d-r)s^{2m}_{-|E_r|,k}(t_{E_0})}{g2^{2m} + (d-r)s^{2m}_{-|E_r|,k}(t_{E_0})}. 
		\end{equation}
		Note that $s_{-|E_r|,k}(t_{E_0})< 2$.

It is easy to extend the proof to work for every $t\in(0,t_{E_0}]$, since all the arguments $t|E_i|/k$ then lie in $[0,\pi/2]$: replace $t_{E_0}$ by $t$, $g\,2^{2m}$ by $g\,s^{2m}_{E_0,k}(t)$, and $2$ by $s_{E_0,k}(t)$ in the bound $s_{-|E_r|,k}<2$. 
	\end{proof}

\section{Proof of limit formula \eqref{eq:filter-floor}}\label{app:filter-floor-proof} 
\begin{proof}
We have for any $t\in [0,+\infty)$
\[	\prod_{p=0}^{m-1}\sin^2(\theta_p)(t)=\frac{\left\|S_k^m(t)|\psi^0\rangle\right\|^2}{2^{2m}}.\]
It is easy to see that for $|\psi^0\rangle= \frac{1}{\sqrt{d}}\sum_{i=0}^{d-1}|E_i\rangle$,
\begin{align*}
	\prod_{p=0}^{m-1}\sin^2(\theta_p)(t)&=\frac{\left\|S_k^m(t)|\psi^0\rangle\right\|^2}{2^{2m}}= \frac{1}{2^{2m}}\sum_{i=0}^{d-1}s_{E_i,m}^{2m}(t)|\langle\psi^0\mid E_i\rangle|^2\\
	&= \frac{1}{2^{2m}d}\sum_{i=0}^{d-1}s_{E_i,m}^{2m}(t),
\end{align*}
In particular, when $t= t_{E_0}= \frac{k \pi}{2 |E_0|}$ 
\[s_{E_i,m}(t_{E_0})=\left\lbrace \begin{array}{lll}
	2&\text{if}& E_i=E_0\\
	\lambda_i\in [0,2)&\text{if}& E_i\neq E_0.
\end{array}\right. \]
Then, 
\begin{align*}
	\prod_{p=0}^{m-1}\sin^2(\theta_p)(t_{E_0})&= \frac{g2^{2m}+\sum_{i=g}^{d-1}\lambda_i^{2m}}{2^{2m}d}= \dfrac{2^{2m}\left( g+\sum_{i=g}^{d-1}(\frac{\lambda_i}{2})^{2m}\right) }{2^{2m}d} \\
	&= \dfrac{g+\sum_{i=g}^{d-1}\left( \frac{\lambda_i}{2}\right) ^{2m} }{d},
\end{align*}
where $g$ denotes the multiplicity of the eigenvalue $E_0$. Therefore, 
\[\lim\limits_{m\to \infty}	\prod_{p=0}^{m-1}\sin^2(\theta_p)(t_{E_0})=\frac{g}{d}=\frac{g}{2^N},\]
since $0\leq \frac{\lambda_i}{2}<1$ and $\lim\limits_{m\to\infty}\left( \frac{\lambda_i}{2}\right) ^{2m}=0$ for every $g\leq i\leq d-1$.
\end{proof}

\section{Total Failure Probability for 20 qubits in sequential QSITE}
\label{app:joint-20q}
  \begin{figure}[H]
    \centering
    \includegraphics[width=0.62\linewidth]{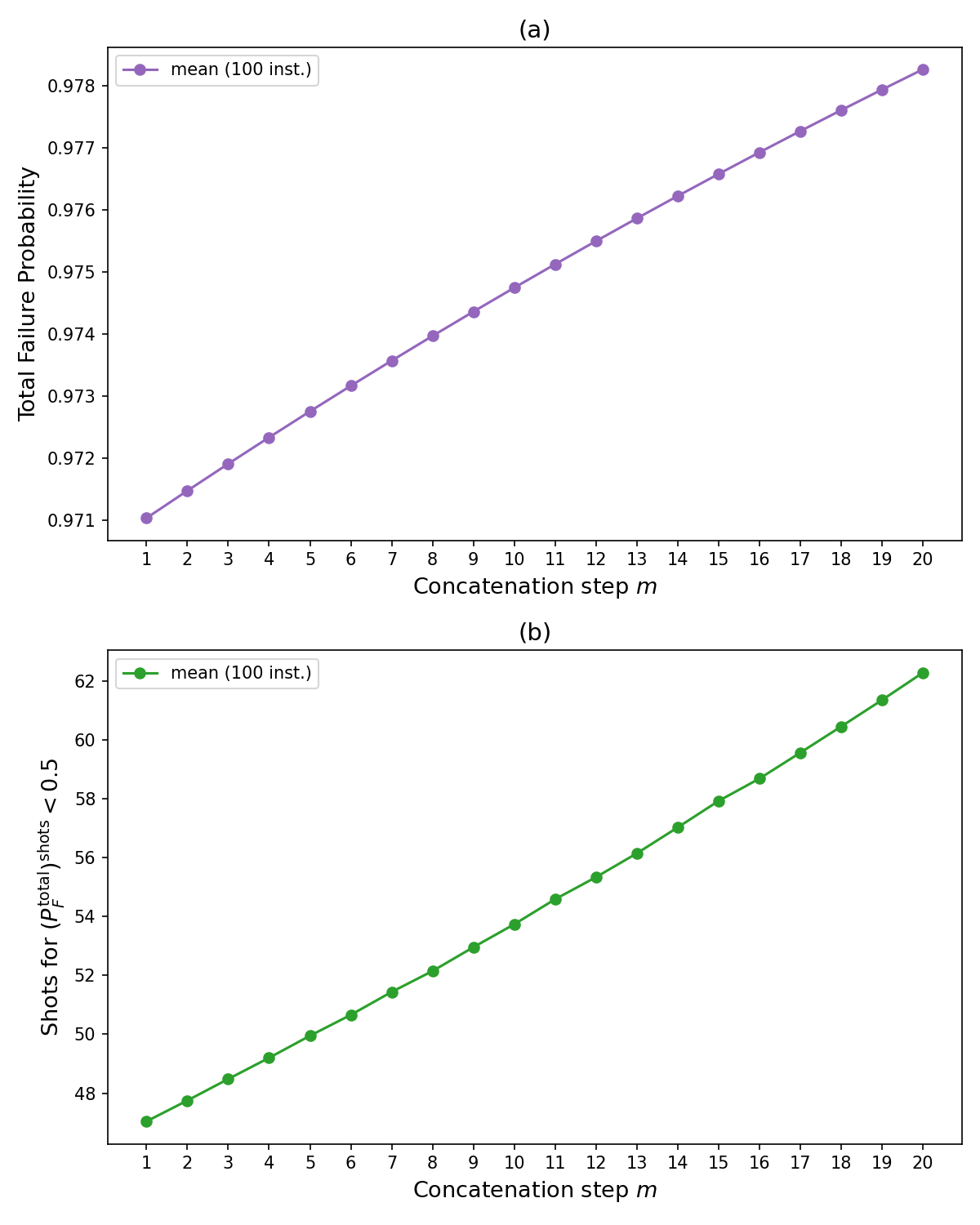}
    \caption{Total failure probability $P_F^\mathrm{total}$ (top) and required number of shots (bottom) as a function of concatenation steps $m$ for a 20-qubit MaxCut ensemble (100 instances, $k=6$, $\delta E = 8 \approx 12\%\,|E_0|$, first safe zone, sequential QSITE). Both quantities increase approximately linearly with $m$, because the decreasing cumulative filter probability slightly outweighs the improving $P_F$.} 
    \label{fig:joint-20q}
\end{figure}

\section{Shots from Probability of Failure Only}
\label{app:shots-pf-only}

  \begin{figure}[H]
    \centering
    \includegraphics[width=0.72\linewidth]{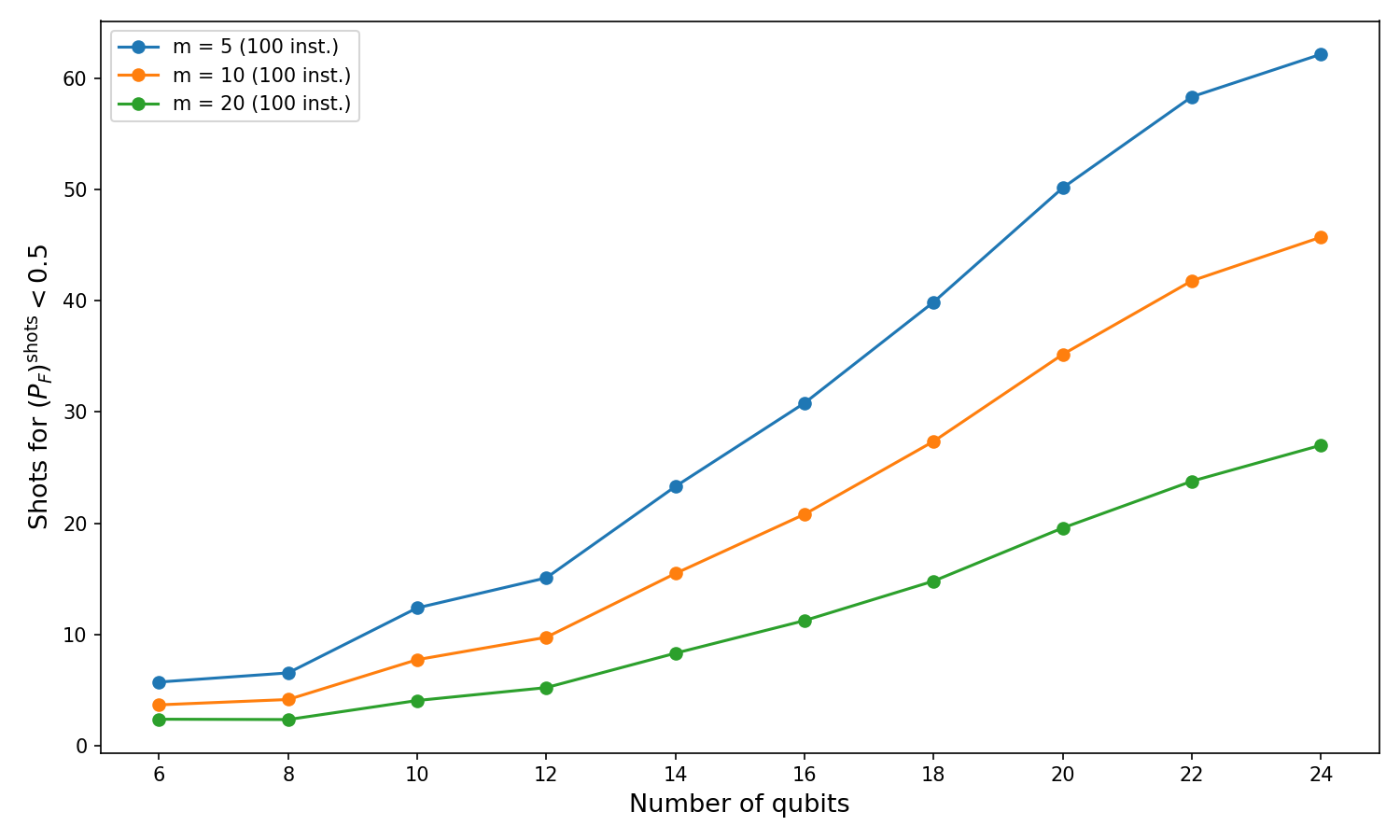}
    \caption{Number of shots $S$ required to achieve $(P_F)^S < 0.5$ as a function of the number of qubits $N$, for $m=5,10,20$ concatenation steps ($k=6$, first safe zone, $\delta E = 12\%\,|E_0|$, sequential QSITE without amplitude amplification). Each point is the mean over 100 random MaxCut instances. Unlike Fig.~\ref{fig:shots-vs-qubits}, filter failures are ignored: shots are computed from $P_F$ alone via $S = \lceil \log(0.5)/\log(P_F) \rceil$. The three curves are clearly separated, confirming that larger $m$ reduces the per-shot $P_F$ and hence the number of repetitions needed.} 
    \label{fig:shots-pf-only}
\end{figure}

\section{Sensitivity to the Energy Tolerance $\delta E$}
\label{app:delta-e-sensitivity}

The number of shots required in the first safe zone (Fig.~\ref{fig:shots-vs-qubits}) is strikingly sensitive to the choice of the energy tolerance $\delta E$. For MaxCut Hamiltonians the eigenvalues are integers, so consecutive energy levels are usually separated by $1$; however, the degeneracy of each level grows rapidly with the energy distance from the ground state. For the 20-qubit instance of Fig.~\ref{fig:20q-Pfsite}, $E_0 = -71$ has degeneracy~$2$, $E_1 = -70$ has degeneracy~$8$, and $E_2 = -69$ has degeneracy~$22$---yet the gap $|E_0 - E_1|$ represents only $1.4\%$ of $|E_0|$ and $|E_0 - E_2|$ only $2.8\%$. Because each additional energy level included within $\delta E$ brings a disproportionately large number of states, a small change in the relative tolerance---from $12\%$ to $15\%$ of $|E_0|$, which shifts $\delta E$ from ${\sim}\,8.5$ to ${\sim}\,10.7$---can include only two extra integer energy levels but nearly four times more states: for this instance, $\delta E = 8.52$ ($12\%$) yields $7{,}922$ states with energy in $[E_0,\, E_0 + \delta E]$, whereas $\delta E = 10.65$ ($15\%$) yields $29{,}742$ out of $2^{20} = 1{,}048{,}576$ total states. 

\begin{figure}[H]
    \centering
    \includegraphics[width=0.72\linewidth]{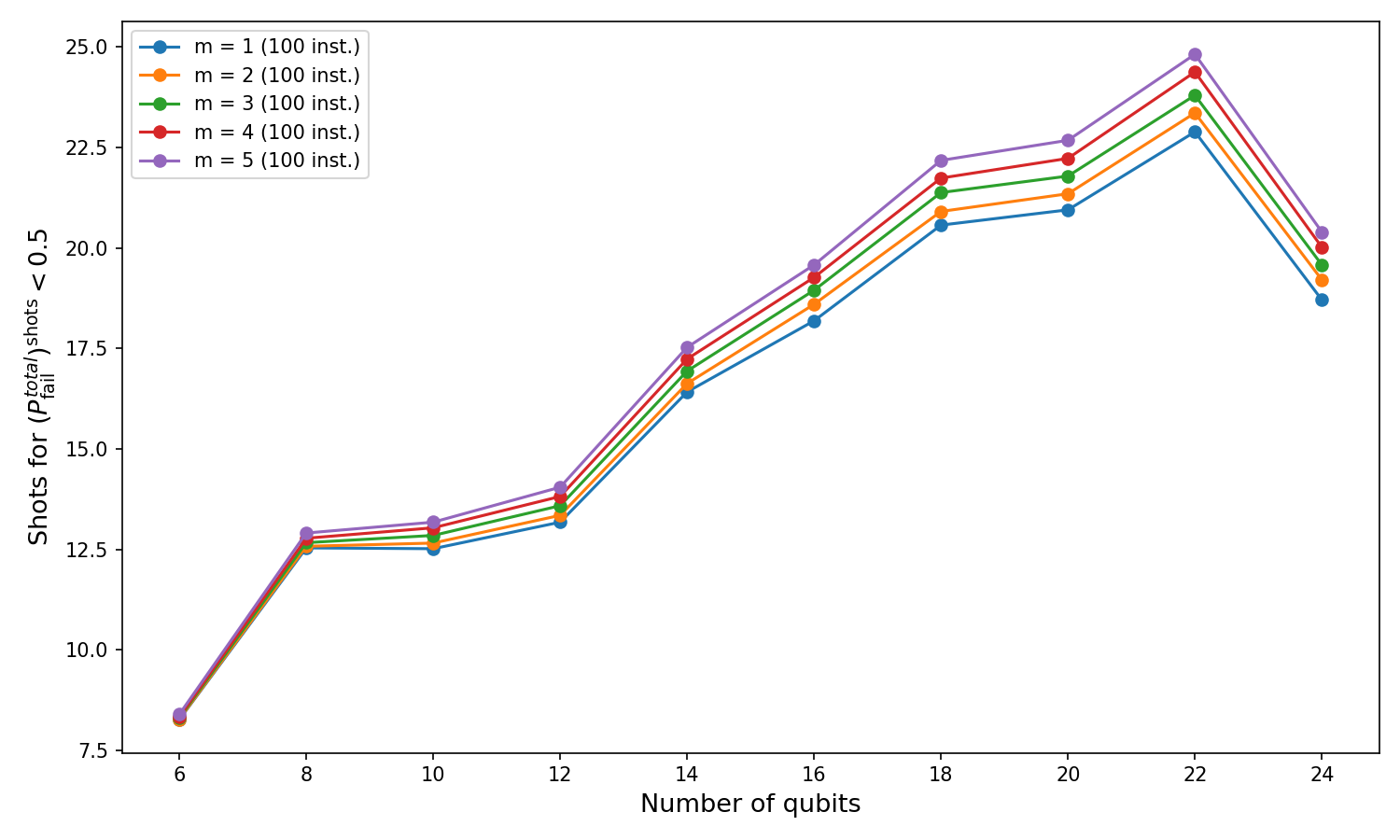}
    \caption{Number of shots required to achieve total failure probability $P_F^\mathrm{total} < 0.5$ as a function of the number of qubits $N$, for $m=1,2,3,4,5$ concatenation steps of the sequential QSITE schedule at the first safe zone ($k=6$, $\delta E = 15\%\,|E_0|$, without amplitude amplification, mean over 100 random MaxCut instances). Compared with the $12\%$ case (Fig.~\ref{fig:shots-vs-qubits}) the shots drop by a factor that grows with $N$, to about $4$ for $N\ge20$.} 
    \label{fig:shots-vs-qubits-15pct}
\end{figure}

Figure~\ref{fig:shots-vs-qubits-15pct} repeats the first-zone analysis of Fig.~\ref{fig:shots-vs-qubits} at the larger tolerance $\delta E = 15\%\,|E_0|$. Compared with the $12\%$ case, the shots drop markedly (reaching only ${\sim}\,25$ at $N = 22$ versus ${\sim}\,90$ at $12\%$), and, as at $12\%$, increasing $m$ leads to a slight \emph{increase} in required shots. A difference of just $3$ percentage points in relative tolerance thus changes the required shots by a factor that grows with $N$, to about $4$ for $N\ge20$; this sensitivity is intrinsic to the combinatorial structure of the spectrum, making $\delta E$ a critical algorithmic parameter. 

\section{Supplementary Shots-Scaling Data}
\label{app:supplementary-shots}

This appendix collects the two shots-scaling results discussed in the main text (Section~\ref{sec:numerical-results}) but not plotted there. 

\begin{figure}[H]
    \centering
    \includegraphics[width=0.72\linewidth]{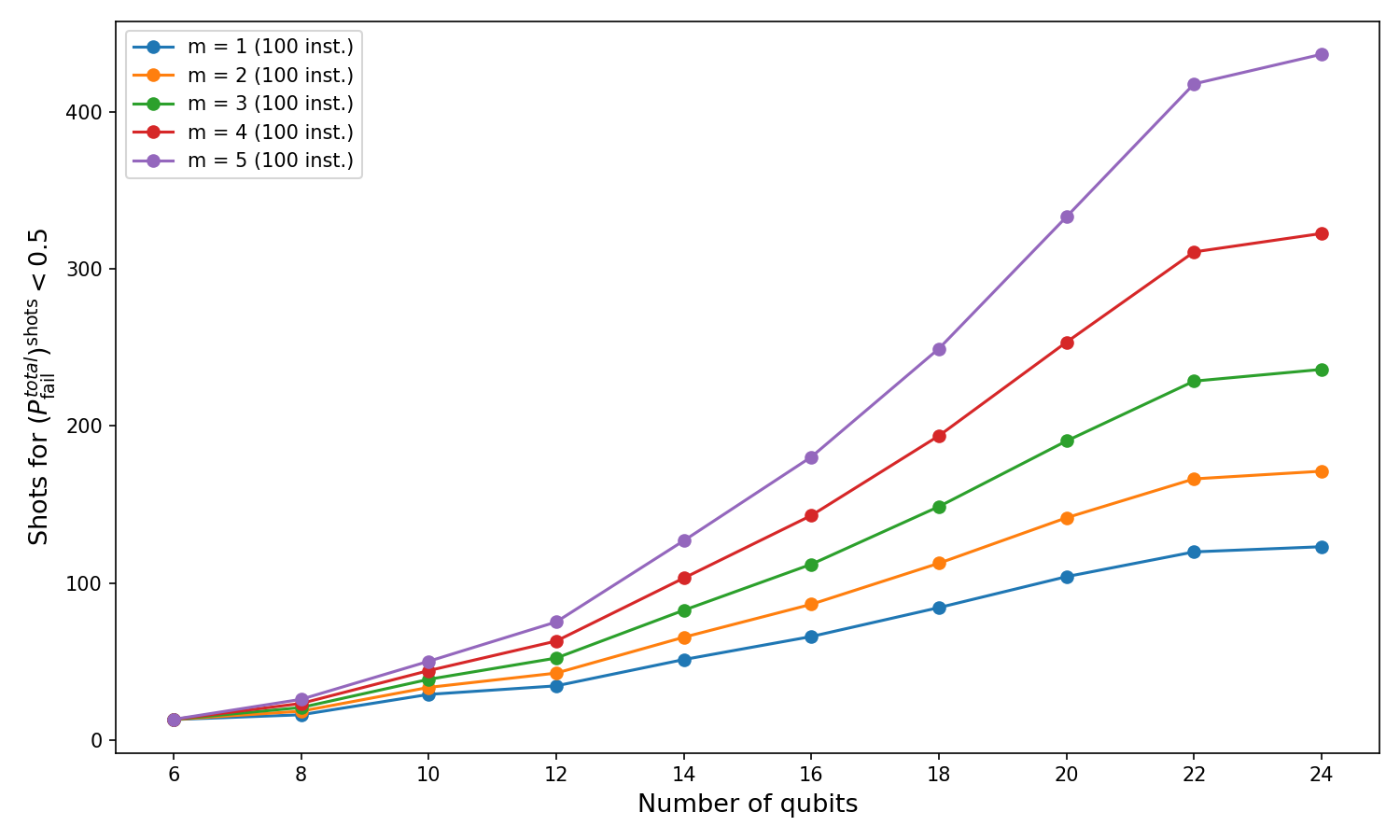}
    \caption{Number of shots required to achieve total failure probability $P_F^\mathrm{total} < 0.5$ as a function of the number of qubits $N$, for $m=1,2,3,4,5$ concatenation steps of the sequential QSITE schedule at the \emph{second} safe zone ($k=6$, $\delta E = 12\%\,|E_0|$, without amplitude amplification, mean over 100 random MaxCut instances). Unlike the first zone (Fig.~\ref{fig:shots-vs-qubits}), the curves separate and the shots grow steeply with $m$, reaching ${\sim}\,437$ at $N=24$ for $m=5$: the smaller cumulative filter probability of the second zone makes deeper concatenation progressively more expensive without amplitude amplification.} 
    \label{fig:shots-vs-qubits-secondzone}
\end{figure}

Figure~\ref{fig:shots-vs-qubits-secondzone} shows the sequential QSITE shots-vs-$N$ analysis at the \emph{second} safe zone, with the same parameters as the first-zone Fig.~\ref{fig:shots-vs-qubits} ($k=6$, $m=1,\dots,5$, $\delta E = 12\%\,|E_0|$, mean over 100 instances). In contrast to the first zone, the curves separate and the required shots grow markedly with $m$, reaching ${\sim}\,437$ at $N=24$ for $m=5$ versus ${\sim}\,123$ for $m=1$ (and versus the roughly $m$-independent ${\sim}\,90$ of the first zone). The much larger shot counts reflect the smaller cumulative filter probability at the second zone, and their growth with $m$ shows that, absent amplitude amplification, deeper concatenation only increases the filter cost without a compensating gain. 

\begin{figure}[H]
    \centering
    \includegraphics[width=0.75\linewidth]{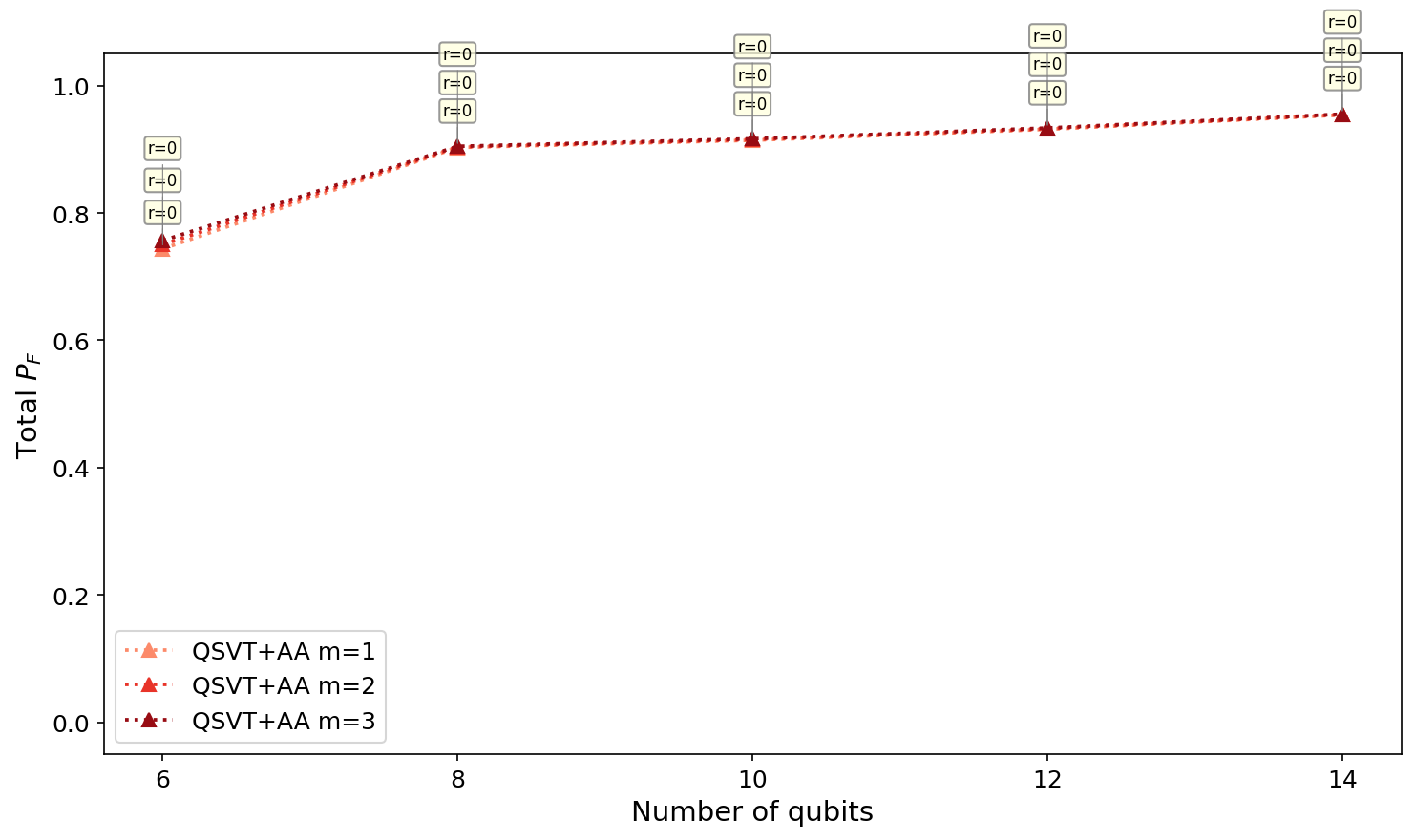} 
    \caption{Per-shot total failure probability $P_F^\mathrm{total}$ of the coherent QSVT realization with Grover amplitude amplification versus number of qubits $N$, at the \emph{first} safe zone ($k=N$, $m=1,2,3$, 100 random MaxCut instances, $\delta E$ around $12\%\,|E_0|$). The mean number of Grover iterations $r=\lfloor \pi/(4\theta)-1/2\rfloor$ is annotated at each point; here $r=0$ throughout, because the first-zone success probability is above $1/4$, leaving amplification inactive. The three degree curves nearly coincide and $P_F^\mathrm{total}$ grows with $N$, reflecting a $P_F$-limited regime at shallow depth. Contrast the second safe zone (Fig.~\ref{fig:qsvt-aa-zone2-vs-qubits}), where amplification is active and deeper polynomials help.} 
    \label{fig:qsvt-aa-zone1-vs-qubits}
\end{figure}

The first-zone counterpart of the coherent QSVT+AA analysis of Fig.~\ref{fig:qsvt-aa-zone2-vs-qubits} is shown in Fig.~\ref{fig:qsvt-aa-zone1-vs-qubits}. With the Grover rule of Sec.~\ref{sec:seq-qsite}, $r=\lfloor \pi/(4\theta)-1/2\rfloor$, we get $r=0$ across the whole range, because the first-zone success probability $\sin^2\theta$ is above $1/4$ for every instance, and for most of them above $1/2$, where even one Grover iteration would lower it. Amplification is therefore inactive and $P_F^\mathrm{total}$ is that of the sequential schedule, which is high and grows with $N$ while the three degree curves nearly coincide. 

\section{Fully Connected Graphs}\label{app:fc}

We now consider fully connected MaxCut graphs (complete graphs $K_N$). For $K_N$ with $N$ even, the MaxCut Hamiltonian has a minimum energy $E_0 = -(N/2)^2$, achieved by all balanced bipartitions of the vertex set. The number of degenerate ground-state configurations is $\binom{N}{N/2}$, which grows exponentially with $N$ (see Table~\ref{tab:fc}).

The QITE algorithm of Motta et al.~\cite{Motta_2019} reproduces each time step by a unitary acting on a domain whose size is set by the correlation length of the evolving state. In $K_N$ all qubits are adjacent and equivalent, and the target state (the uniform superposition of balanced bitstrings) has identical correlations between every pair of qubits, so exact QITE requires domains spanning all $N$ qubits, at a cost exponential in $N$; smaller domains can only be used heuristically. QSITE instead uses the same two-ancilla circuit for any connectivity, with $O(|E|)=O(N^2)$ gates per step for $K_N$. 

\begin{table}[H]
\centering
\caption{Properties of fully connected MaxCut instances $K_N$ for even $N$ from 6 to 24. $|E| = N(N-1)/2$ denotes the number of edges, $E_0 = -(N/2)^2$ the ground-state energy, and $g = \binom{N}{N/2}$ its degeneracy.}
\label{tab:fc}
\begin{ruledtabular}
\begin{tabular}{cccc}
$N$ & $|E|$ & $E_0$ & $g = \binom{N}{N/2}$ \\
\hline
6  & 15  & $-9$   & 20          \\
8  & 28  & $-16$  & 70          \\
10 & 45  & $-25$  & 252         \\
12 & 66  & $-36$  & 924         \\
14 & 91  & $-49$  & 3\,432      \\
16 & 120 & $-64$  & 12\,870     \\
18 & 153 & $-81$  & 48\,620     \\
20 & 190 & $-100$ & 184\,756    \\
22 & 231 & $-121$ & 705\,432    \\
24 & 276 & $-144$ & 2\,704\,156 \\
\end{tabular}
\end{ruledtabular}
\end{table}

\begin{figure}[H]
    \centering
    \includegraphics[width=0.85\linewidth]{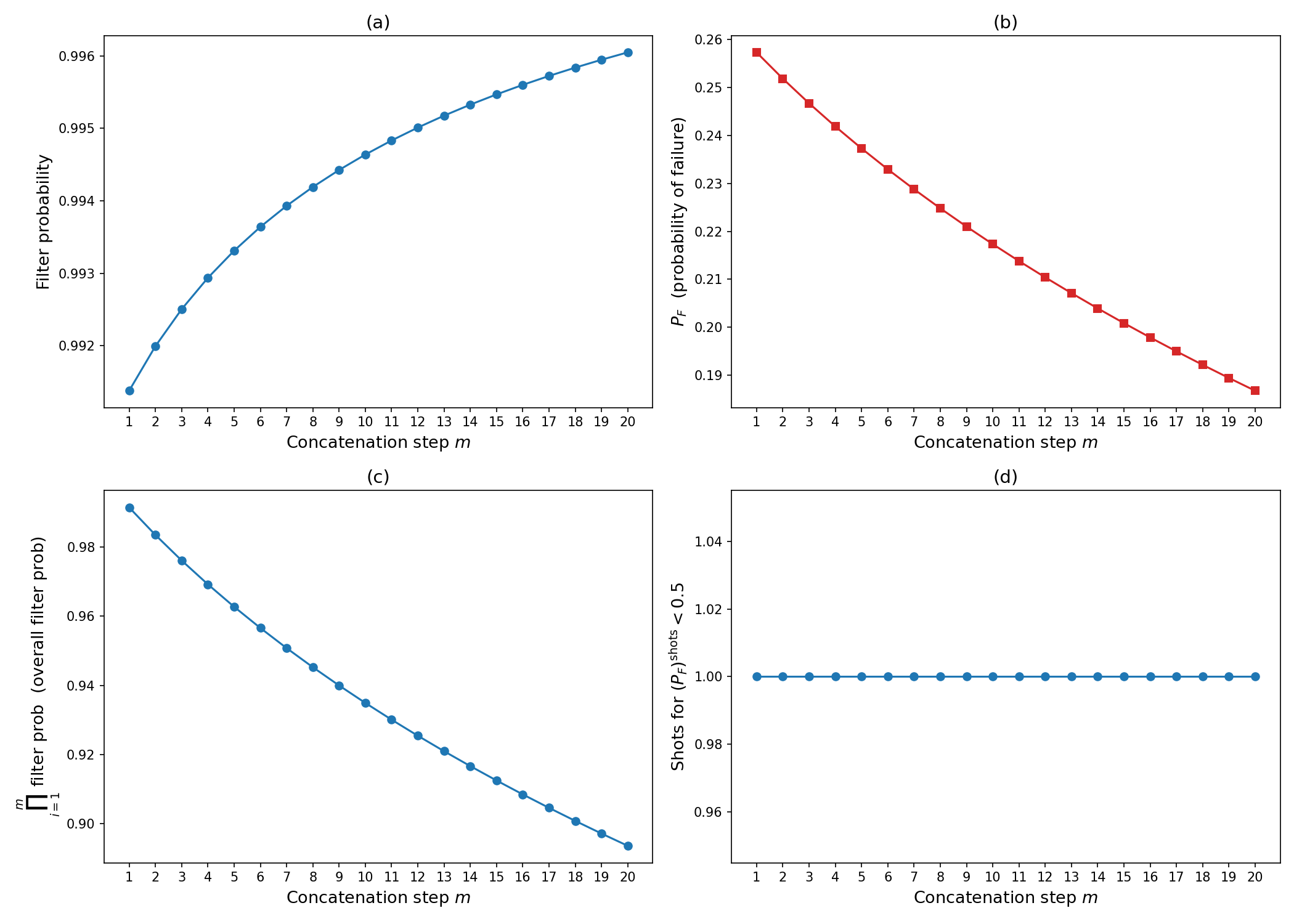}
    \caption{Filter probability (a), failure probability $P_F$ (b), cumulative filter probability (c), and number of shots $S$ with $(P_F)^S<0.5$, ignoring filter failures (d), for a fully connected 20-qubit MaxCut instance ($K_{20}$, $E_0=-100$, degeneracy $\binom{20}{10}=184{,}756$, $\delta E = 8$, $k=6$) evaluated at the first safe zone. Fewer than 2 shots are required throughout, reflecting the high ground-state degeneracy.} 
    \label{fig:fc-firstzone}
\end{figure}

Figure~\ref{fig:fc-firstzone} presents the four-panel analysis for a 20-qubit fully connected graph ($K_{20}$) at the first safe zone. The failure probability decreases with concatenation steps and the number of required shots is low: fewer than 2 shots suffice to achieve $P_F^\mathrm{total} < 0.5$, even without amplitude amplification. This is a consequence of the high degeneracy: with $\binom{20}{10} = 184{,}756$ degenerate ground states among $2^{20} \approx 10^6$ total basis states, approximately $17.6\%$ of all computational basis states are ground-state configurations. 

\begin{figure}[H]
    \centering
    \includegraphics[width=0.62\linewidth]{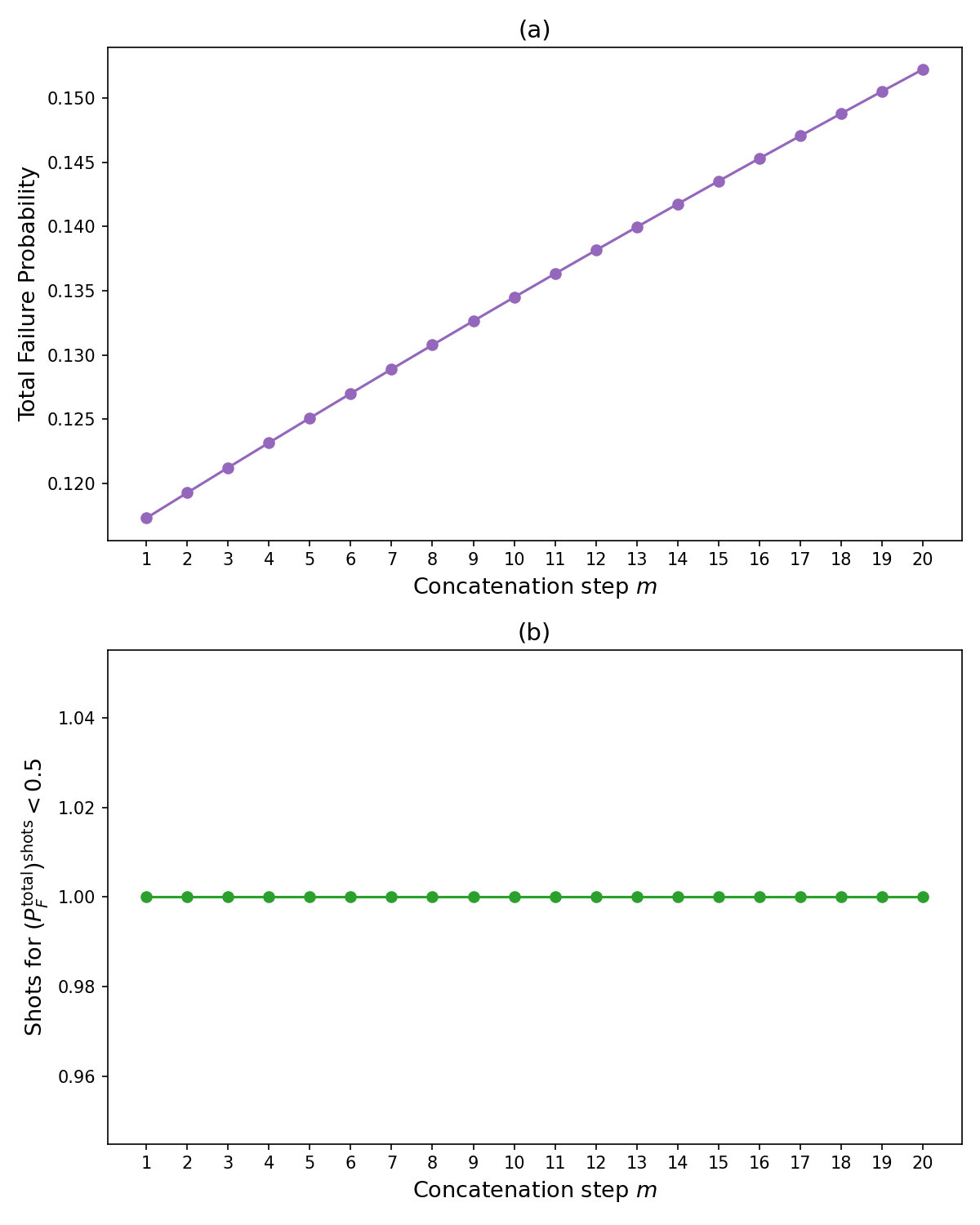}
    \caption{Total failure probability $P_F^\mathrm{total}$ (a) and required number of shots (b) as a function of concatenation steps $m$ for the fully connected 20-qubit MaxCut instance ($K_{20}$, $E_0=-100$, $\delta E = 10\%\,|E_0|$, $k=6$, first safe zone, sequential QSITE). $P_F^\mathrm{total}$ grows linearly but stays well below $0.5$, so only 1 shot is needed at every concatenation depth.} 
    \label{fig:fc-joint}
\end{figure}

Figure~\ref{fig:fc-joint} shows the total failure probability and the corresponding shots as a function of concatenation steps $m$ (up to $m=20$) for the same $K_{20}$ instance. The total failure probability increases approximately linearly with $m$, consistent with the near-cancellation observed for random graphs, but remains below $0.16$ even at $m=20$. 

\bibliographystyle{ieeetr}
\bibliography{Referencias}

\end{document}